\documentclass[usenatbib]{mn2e}
\usepackage{graphicx}
\usepackage{amsbsy}
\usepackage{amssymb}
\usepackage{color}

\newcommand{\HL}[1]{{\color{black}{#1}}}

\makeatletter
\let\oldcr=\@tabularcr
\makeatother

\title[On the seismic age and heavy-element abundance of the Sun]
      {On the seismic age and heavy-element abundance of the Sun}
\author[G. Houdek \& D. O. Gough]
       {G. Houdek$^{1}$\thanks{E-mail: guenter.houdek@univie.ac.at},
        D. O. Gough$^{2}$\thanks{E-mail: douglas@ast.cam.ac.uk}\\
        $^{1}$Institute of Astronomy, University of Vienna, 
              1180 Vienna, Austria\\
        $^{2}$Institute of Astronomy and Department of Applied Mathematics
              and Theoretical Physics, \\
              University of Cambridge, Cambridge CB3 0HA, UK
       }
\begin{document}

\date{Accepted 2011 August 2. Received 2011 August 2; 
      in original form 2011 April 29}

\maketitle

\label{firstpage}

\begin{abstract}
We estimate the main-sequence age and heavy-element abundance
of the Sun by means of an asteroseismic calibration of theoretical
solar models using only low-degree acoustic modes from 
the BiSON. The method can therefore be applied also to other 
solar-type stars, such as those observed by the NASA satellite 
Kepler and the planned ground-based Danish\HL{-led} SONG network.
The age, 4.60$\pm$0.04 Gy, obtained with this new seismic method, 
is similar to, \HL{although somewhat greater than,} today's commonly 
adopted values, and the surface 
heavy-element abundance by mass, $Z_{\rm s}$=0.0142$\pm0.0005$, lies
between the values quoted recently by Asplund et al. (2009) and by
Caffau et al. (2009). We stress that our best-fitting model is not
a seismic model, but a theoretically evolved model of the Sun
constructed with `standard' physics and calibrated against helioseismic data.
\end{abstract}


  \section{INTRODUCTION}
  \label{sec:intro}

The only way by which the age of the Sun can be estimated directly 
to a useful degree
of precision is by accepting the basic tenets of solar-evolution theory and
measuring those aspects of the structure of the Sun that are predicted by
the theory to be indicators of age. 
\HL{We recognize that there are also indirect methods based on the more 
reliable determination of the ages of meteorites 
\citep[e.g.][]{amelin02, jacobsen08, bouvier10}.}
We recognize also that there is not a precise 
\HL{origin of time, such as a}
moment that one can uniquely 
define to be the time at which the Sun arrived on the main sequence. However, 
after initial transients, the central hydrogen abundance $X_{\rm c}$ declined
almost linearly with time \citep[e.g.][]{dog95}, so one can extrapolate 
$X_{\rm c}(t)$ backwards quite well to the time when $X_{\rm c}=X_0$, the
initial hydrogen abundance. That is the time that we adopt as our 
fiducial origin.
\HL{A potential goal of future investigations of the
type we describe here could be to ascertain whether the Sun arrived on the
main sequence before the rest of the solar system formed, or at the same
time. Unfortunately we have not yet succeeded in resolving the matter, partly because
the data errors are not yet small enough, but mainly, as we discuss in
\S\,4, because the uncertainties in the modelling are too great.}

The solar structure measurements must be
carried out seismologically, and one is likely to expect greatest reliability
of the results when all the available pertinent helioseismic data are employed.
Of these, the most pertinent are the frequencies of the modes of lowest degree, 
because it is they that penetrate the most deeply into the energy-generating 
core where the helium-abundance variation records the integrated history of 
nuclear transmutation. Moreover, it is also only they that can be measured in 
other stars. Therefore, there has been some interest in calibrating
theoretical stellar models using only low-degree modes -- here we use modes of
degrees $l$=0, 1, 2 and 3. 
The prospect was first discussed in detail by 
\citet{jcd84, jcd88}, \cite{ulrich86} 
and \cite{dog87}, although prior to that it had already been pointed out
that the helioseismic frequency data that were available at the time 
indicated that either the initial helium abundance $Y_0$, or the age $t_\odot$,
or both, are somewhat greater than the generally accepted values 
\citep[\citealt{dog83}; see also][]{gk90}. Subsequent, more careful, calibrations 
were discussed by \cite{guenther89}, \cite{gn90}, \cite{guenther-demarque97}, 
\cite{weiss-schlattl98}, \cite{wd99}, \cite{dog01}, \citet{bsp02} and
\citet{dbc11}; \HL{all but the last have been reviewed by \cite{jcd09}, who dicusses 
some of the obstacles that need to be surmounted.}
Most of \HL{the calibrations} did not address the influence of uncertainties
in chemical composition on the determination of $t_\odot$;
for example, \cite{weiss-schlattl98} adopted in their calibration the
helioseismically determined values for the helium abundance
in the convection zone,
\HL{together with the convection-zone depth.}

As a main-sequence star ages, helium is produced in the core, increasing
the mean molecular mass $\mu$ preferentially near the centre, and thereby
inducing a local positive gradient of the sound speed. The resulting 
functional form of the sound speed
$c(r)$ depends not only on age $t_\odot$ but also on the relative
augmentation of $\mu(r)$, which itself depends on the initial absolute
value of $\mu$, and hence 
\HL{on the chemical composition:  directly on the initial helium abundance $Y_0$, 
via the equation of state,} 
and, to a lesser degree, $Z_0$, 
\HL{and indirectly, via the model calibration to the observed 
values ${\rm R}_\odot$ and ${\rm L}_\odot $ of the radius $R$ and the 
luminosity $L$, on $Z_0$ and, to a lesser degree, $Y_0$.} 
\cite{dog01} tried to separate these 
two dependencies using the degree dependence of the small separation
$d_{n,l}=3(2l+3)^{-1}(\nu_{n,l}-\nu_{n-1,l+2})$ between cyclic multiplet
frequencies $\nu_{n,l}$, where $n$ is order and $l$ is degree. 
This is possible, in
principle, because modes of different degree and similar frequency sample the
core differently. However, the difference 
between the effects of $t_\odot$ and
$Y_0$ on the functional form of $c(r)$ in the core is not very great, and
consequently the error in the calibration produced by errors in the observed
frequency data is uncomfortably high,
\HL{as is also the case when a mean value of the large separation
$\langle\nu_{n,l}-\nu_{n-1,l}\rangle$ is used in conjunction with the 
mean small separation \citep[][]{gn90}.}

This lack of sensitivity can be overcome by using, in addition to 
core-sensitive seismic signatures, the relatively small oscillatory component 
of the eigenfrequencies induced by the sound-speed glitch associated with
helium ionization \citep{dog02}, whose amplitude is close to
being proportional to helium abundance $Y$ 
\citep{hg07b}. The neglect of that component
in the previously employed asymptotic signature had not only omitted
an important diagnostic of $Y$, but had 
\HL{appeared to}
imprint an oscillatory
contamination in the calibration as the limits $(k_1, k_2)$, where
$k=n+\frac{1}{2}l$, of the adopted mode range was varied 
\citep{dog01}. It therefore behoves us to decontaminate the core
signature from glitch contributions produced in the outer layers of the star
(from both helium ionization and the abrupt variation at the base of the 
convection zone, and also from hydrogen ionization and the superadiabatic
convective boundary layer immediately beneath the photosphere). To this end
a helioseismic glitch signature has been developed by
\cite{hg07b}, from which its contributions $\delta\nu_{n,l}$ to the frequencies
can be computed and subtracted from the raw frequencies
$\nu_{n,l}$ to produce effective glitch-free frequencies $\nu_{{\rm s}n,l}$
to which a glitch-free asymptotic formula -- equation (\ref{e:asymp}) -- can 
be fitted.
The solar calibration is then accomplished as previously \citep{dog01} by 
fitting theoretical seismic signatures to the observations by Newton-Raphson 
iteration, using a carefully computed grid of calibrated models to compute 
derivatives with respect to  \HL{$Z_0$ and the 
age $t_\star$ of each model}.  The result of the 
first preliminary calibration by this method, using
BiSON data, has been reported by \cite{hg07a}. 
Here we enlarge on our discussion of the analysis, taking a more consistent
account of the surface layers of the star, augmenting the number of diagnostic
frequency combinations used in the calibration, and adding a second
starting reference solar model to demonstrate the insensitivity of the
iterated solution to starting conditions. 
\HL{We fit our model of the
frequencies to the BiSON data discussed by \citet{basu07}: they
are mean frequencies obtained over the 4752 days from 1992 December 31
to 2006 January 3 of modes of degree $l=0$, 1, 2, and 3, adjusted to take some account 
of solar-cycle variation}


\begin{figure}
\centering
\includegraphics[width=1.00\linewidth]{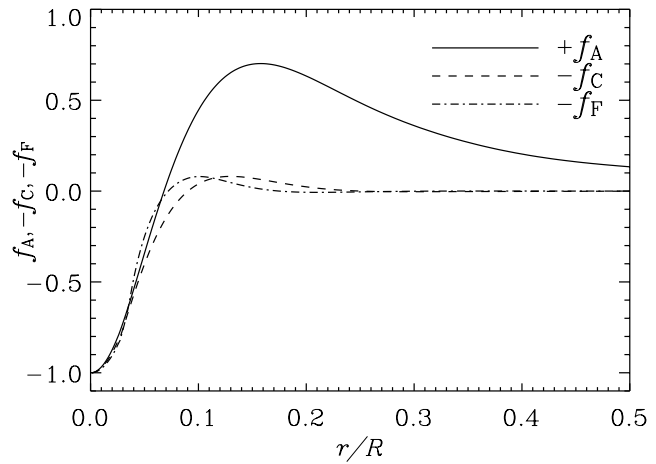}
\caption{
   Functional forms $f_X$ of the integrands $\phi_X$ in 
   $X=\int_0^R \phi_X {\rm d}r$, where 
   \HL{$f_X(r) = \phi_X(r)/|\phi_X(0)|$ }
   and where $X =$\, $A$, $C$, or $F$, plotted for Model S of \citet{jcd96}
   over the inner half of 
   the interval $(0,R)$ of $r$. The parameters $A$, $C$ and $F$ are 
   sensitive particularly to the structure of the core, being progressively 
   more centrally concentrated.   
} 
\label{f:integrals}
\end{figure}


   \section{The calibration procedure}
   \label{sec:calibproc}

   \subsection{Introductory remarks}
   \label{sec:introremarks}
Naively fitting eigenfrequencies of parametrized solar models to observed solar
oscillation frequencies is temptingly straightforward, and was one of the 
earliest procedures to be adopted in the present context \citep{cdg81}.
However, it is unwise to adopt so crude a strategy because the raw frequencies are
affected by properties of the Sun that are not directly pertinent to the
particular investigation in hand, as was quickly realized at the time
\citep[e.g.][]{dog83, cdg84}. An example is the effect of the near-surface 
layers, unwanted here, yet a serious contaminant because the region is one of
low sound speed. It is more prudent to design seismic diagnostics that are
sensitive only to salient properties of the structure. This we accomplish by
noticing the roles of various structural features in asymptotic analysis, and
relating functionals arising in that analysis to corresponding combinations
(not necessarily linear) of oscillation frequencies. It is these combinations
that are then used for the calibration.

We emphasize that the calibration is carried out by processing numerically
computed eigenfrequency diagnostics in precisely the same manner as the observed 
frequencies. After the diagnostics have been designed, asymptotics play no further 
role. The precision of the calibration
itself is independent of the accuracy of the asymptotic analysis; it is only
the accuracy of the conclusions drawn from these calibrations that \HL{is} so
reliant, for those conclusions depend in part on the degree to which the diagnostic 
quantities of, in our present study, age and heavy-element abundance, are 
divorced from extraneous influences. 

   \subsection{Diagnosis of the smoothed structure}
   \label{sec:diagnosis}
The principal age-sensitive diagnostics are contained in the asymptotic 
expression 
\begin{eqnarray}
\nu_{{\rm s}{\boldsymbol i}}\!&\sim&\!(n+{\textstyle\frac{1}{2}}\,L+\varepsilon)\nu_0
-\frac{AL^2\!\!-\!B}{\nu_{{\rm s}{\boldsymbol i}}}\,\nu^2_0
-\frac{CL^4\!\!-\!DL^2\!+\!E}{\nu_{{\rm s}{\boldsymbol i}}^3}\,\nu^4_0\cr
&&-\frac{FL^6\!\!-\!GL^4\!+\!HL^2\!-\!I}{\nu_{{\rm s}{\boldsymbol i}}^5}\,\nu^6_0
=:S_{\boldsymbol i}\,,
\label{e:asymp}
\end{eqnarray}
in which ${\boldsymbol i}=(n,l)$ labels the mode, $L=l+1/2$, and the coefficients 
$\xi_\beta:=(\nu_0, \varepsilon, A, B, ..., I)$, $\beta=1,...,11$, are
functionals of the solar structure alone, independent of ${\boldsymbol i}$. This
formula can be obtained by expanding in inverse powers of frequency the
coupled pair of second-order differential equations 
governing the linearized adiabatic oscillations of a spherically symmetric 
star, as did \citet{tass80}, and at each order solving the resulting equation-pairs 
successively in JWKB \citep{g07} approximation. Alternatively, perhaps more 
conveniently, but maybe less accurately, one can adopt an approximate 
second-order equation 
\HL{(which takes into account the perturbed gravitational potential only partially)
and expand it alone in the limit $n/L\rightarrow\infty$
\citep[e.g.][]{g86b, dog93}.}
The formula~(\ref{e:asymp}) approximates the actual (adiabatic) eigenfrequencies, 
for finite $n$, only if the scale $H$ of variation of the background 
equilibrium state is everywhere much greater than the inverse vertical wavenumber 
of the oscillation mode. That is accomplished by regarding the solar model, 
$\cal M$, to have been replaced by a smooth model, ${\cal M}_{\rm s}$, from which
the acoustic glitches have been removed. We denote its frequencies by 
$\nu_{{\rm s}{\boldsymbol i}}$.

The coefficients in expression~(\ref{e:asymp}) that are most
sensitive to the stratification of the core are those multiplying the highest
powers of $L$ at each order in $\nu_0/\nu_{{\rm s}{\boldsymbol i}}$, namely 
$A,\ C,$ and $F$. (The $L$-dependent part of the leading
term is also sensitive to the core, but merely to indicate, in the spherical 
environment, that there is no seismically detectable physical singularity at the 
centre of the star; there is, of course, a coordinate singularity in spherical polar coordinates.)
The next terms in core sensitivity are $D$ and $G$, and then $H$. These are also
sensitive to the structure of the envelope, so
we ignore them in the calibration. Below the near-surface layers of a spherically 
symmetrical star the integrands for $A,\ C$ and $F$ (which here we denote by the parameter
$\alpha=1, 2, 3$ respectively) are given approximately by 
\HL{
\begin{equation}
\frac{(-1)^\alpha}{(2\alpha-1)2^\alpha\alpha!}
\left(\frac{1}{r}\frac{{\rm d}}{{\rm d}r}\right)^\alpha
\left(\frac{c}{\nu_0}\right)^{2\alpha-1}
\label{e:integrands}
\end{equation}
}
\noindent \citep{g11}, where $r$ is a radial co-ordinate 
\HL{and $c$ is the adiabatic sound speed;} they are plotted
in Fig.\,\ref{f:integrals}. Notice that the higher the order in the expansion,
the more concentrated near the centre of the star is the integrand of the most
sensitive functional. The integrands depend on progressively higher derivatives of 
the sound speed. Moreover their evaluation \HL{by fitting formula (1) to oscillation 
frequencies}  is more susceptible to frequency errors. 
Granted that we use frequencies of modes of only
four different degrees, $l$=0, 1, 2 and 3,  we cannot even in principle 
determine 
\HL{from them}
coefficients arising in terms of higher order than those presented 
in the truncated expansion~(\ref{e:asymp}).

One can see from expression\,(\ref{e:integrands}) for the
integrands of the coefficients $A$, $C$, and $F$ that they depend 
also on $\nu_0$, which is sensitive
to the outer layers of the star where the sound speed is low. We remove 
that sensitivity by eliminating
$\nu_0$ from expression\,(\ref{e:integrands}), and using instead for our diagnostics the parameters
$\hat A=\nu_0A,\ \hat C=\nu_0^3C$ and $\hat F=\nu_0^5 F$,
\HL{which are the natural factors arising in the asymptotic expansion~(\ref{e:asymp})  of 
$\nu_{{\rm s}{\boldsymbol i}}$ in inverse powers of $\nu_{{\rm s}{\boldsymbol i}}/\nu_0$}.

   \subsection{Glitch contributions}
   \label{sec:glitchcont}
The abrupt variation in the stratification of a star (relative to the scale of 
the inverse radial wavenumber of a seismic mode of oscillation), associated
with the depression in the first adiabatic exponent 
$\gamma_1=(\partial {\ln p}/\partial{\ln\rho})_s$ (where $p$, $\rho$ and $s$ 
are pressure, density and specific entropy) caused by helium ionization, 
imparts a glitch in the sound speed $c(r)$, which induces an oscillatory 
component in the spacing of the eigenfrequencies of low-degree seismic modes
\citep{g90a}. 
The amplitude of the oscillations is an increasing function of the helium 
abundance $Y$, and, for a given adiabatic `constant' $p/\rho^{\gamma_1}$, 
is very nearly proportional to it \citep{hg07b}.
It is therefore a good diagnostic of $Y$. To determine the amplitude we
construct a deviant
\begin{equation}
\delta\nu_{\boldsymbol i}:=\nu_{\boldsymbol i} - \nu_{{\rm s}{\boldsymbol i}}\,
\label{e:nudiff}
\end{equation}
from the frequency $\nu_{{\rm s}{\boldsymbol i}}$ 
of a similar smoothly stratified star, presuming that $\nu_{{\rm s}{\boldsymbol i}}$
is described approximately by equation~(\ref{e:asymp}).

\begin{figure}
\centering
\includegraphics[width=0.9\linewidth]{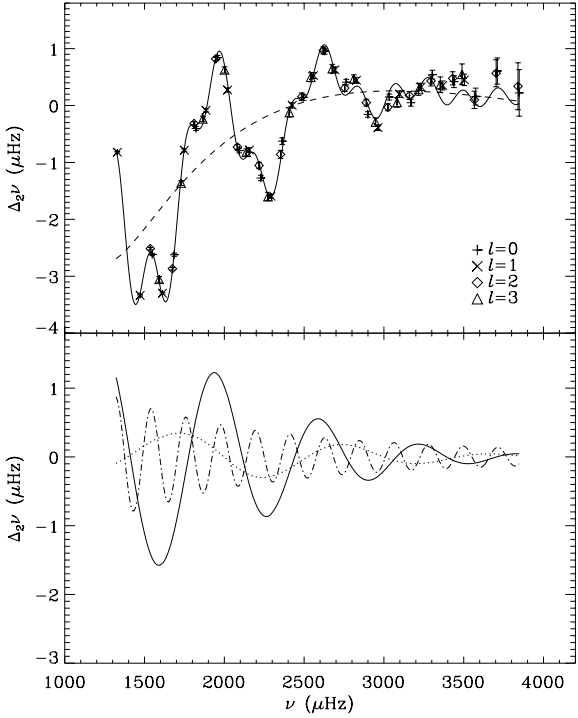}
\caption{
          The symbols in the {\bf upper panels} denote second differences
          $\Delta_{2{\boldsymbol i}}\nu:=\nu_{n-1,l}-2\nu_{n,l}+\nu_{n+1,l}$ of low-degree 
          modes obtained from the BiSON (Basu et al. 2007). The solid curve 
          is a fit of the seismic diagnostic (equation\,\ref{e:all_glitches})
          to the data by appropriately weighted least squares. The 
          dashed curve is the smooth contribution, including a third-order 
          polynomial in $\nu^{-1}_{\boldsymbol i}$ to represent the upper-glitch 
          contribution from near-surface effects 
      \HL{and a contribution from the 
          (leading-order) second differences of $\nu_{{\rm s}{\boldsymbol i}}$ 
          given by equation\,(\ref{e:asymp}), as descibed in the text (\S\,2.3).  }
          The {\bf lower panel} displays the 
          remaining individual oscillatory contributions (with zero means) from 
          the acoustic glitches to $\Delta_{2{\boldsymbol i}}\nu$: the dotted and 
          solid curves are the contributions 
          from the first and second stages of helium ionization, and the 
          dot-dashed curve is the contribution from the acoustic glitch at the 
          base of the convective envelope.}
\label{f:BiSON}
\end{figure}

A convenient and easily executed procedure for estimating the amplitude of the
oscillatory component is via the second multiplet-frequency difference 
with respect to order $n$ amongst modes of like degree $l$:
\begin{equation}
\Delta_2\nu_{\boldsymbol i}:=\nu_{n-1,l}-2\nu_{n,l}+\nu_{n+1,l}\,.
\label{e:secdiff}
\end{equation}
Taking such a difference suppresses smoothly
varying (with respect to $n$) components. The oscillatory component in
$\Delta_2\nu$, produced by an acoustic glitch, has a `cyclic frequency' 
approximately equal to twice the acoustic depth
\begin{equation}
\tau=\int_{r}^R c^{-1}\,{\rm d}r
\end{equation}
of the glitch. The amplitude depends on the amplitude $\Gamma$ and radial
extent $\Delta$ of the glitch, and decays with $\nu$ once the inverse radial 
wavenumber of the mode becomes comparable with or less than $\Delta$.

The effects on the frequencies of a solar model $\cal M$ of a specific glitch
perturbation $\delta\gamma_1$ can most readily be estimated from a variational
principle in the form $\nu={\cal K}/{\cal I}$, as have \citet{g90a}, 
\HL{\citet{h04}, \citet{hg04}} 
and \citet{mt05}. 
\citet{hg07b} have found that a good approximation to the outcome is
\begin{equation}
\delta_\gamma\nu=\frac{\delta_\gamma{\cal K}}{8\pi^2{\cal I}\nu}\,,
\label{e:vp}
\end{equation}
where
\begin{equation}
{\cal I}:=\int\rho\;\bxi\!\cdot\!\bxi\;r^2{\,\rm d}r
\label{e:inertia}
\end{equation}
is the mode inertia and
\begin{equation}
\delta_\gamma{\cal K}\simeq
    \int\delta\gamma_1\;p({\rm div}\bxi)^2r^2\,{\rm d}r\,.
\label{e:K1}
\end{equation}
The function $\bxi$ is the displacement eigenfunction associated with
either $\cal M$ or a corresponding smooth model; here we implicitly use
${\cal M}_{\rm s}$. Several terms in equations (\ref{e:vp}) and (\ref{e:K1}) 
are missing from the exactly perturbed equation; these are relatively small, 
and in any case to a substantial degree they cancel.

The next step of the estimation is to select a convenient representation for
$\delta\gamma_1$. Several formulae have been suggested and used, by e.g., 
\citet{mt98, mt05}, \citet{b04}, \citet{bm04}, and \citet{dog02}, not all of which
are derived directly from explicit acoustic glitches representing helium ionization
\citep[e.g.][]{b97}.  Gough used a single Gaussian function; in 
contrast, Monteiro \& Thompson assumed a triangular form;
Basu et al. adopted a simple discontinuity 
\HL{\citep[][]{ban94}.}
The artificial discontinuities 
in the sound speed and its derivatives that the latter two possess cause 
the amplitude of the oscillatory signal to decay with frequency too gradually,
although that deficiency may not be immediately noticeable within the limited 
frequency range in which adequate asteroseismic data are or will 
imminently be available.  
\HL{The analytic representation, namely a Gaussian function, 
which was used by \citet{dog02} and \citet{h04}, can be made to fit the glitch 
frequency perturbation more closely, especially if the frequency range 
is large. 

All these early representations addressed only the second 
stage of helium ionization.}
Subsequently 
\HL{\citet{hg04, hg06, hg07a, hg07b}}
added another Gaussian function to take account of 
the first stage of helium ionization, relating its location, $\tau_{\rm I}$, 
amplitude factor, $\Gamma_{\rm I}$, and width, $\Delta_{\rm I}$, to those
of the second stage according to a standard 
solar model; and thereby they attained considerable improvement.
Accordingly, we adopt that procedure here, and set
\begin{equation}
\frac{\delta\gamma_1}{\gamma_1}=-\frac{1}{\sqrt{2\pi}}\sum_{i=1}^2
\frac{\Gamma_i}{\Delta_i}
{\rm e}^{-(\tau-\tau_i)^2/2{\Delta^2_i}}\,,
\label{eq:dgog}
\end{equation}
summing over the two stages $i$ (=$\,$I and II) of ionization. We set
$\Gamma_{\rm I}\Delta_{\rm I}/\Gamma_{\rm II}\Delta_{\rm II} = \tilde\beta$,
$\tau_{\rm I}/\tau_{\rm II}=\tilde\eta$, and
$\Delta_{\rm I}/\Delta_{\rm II}=\tilde\mu$.
We have found that $\tilde\beta, \tilde\nu$ and $\tilde\mu$ hardly vary as $Y_0$
and $t_\odot$ are varied in calibrated solar models, and we set their values to
be the constant values 0.45, 0.70, and 0.90 respectively, which gives the best
fit \citep{hg07b}. 
The quantities $\tau_{\rm II}, \Gamma_{\rm II}$ and $\Delta_{\rm II}$, or 
equivalently   $\tau_{\rm I }, \Gamma_{\rm I }$ and $\Delta_{\rm I }$, are
adjustable parameters of the calibration.

\begin{figure}
\centering
\includegraphics[width=1.0\linewidth]{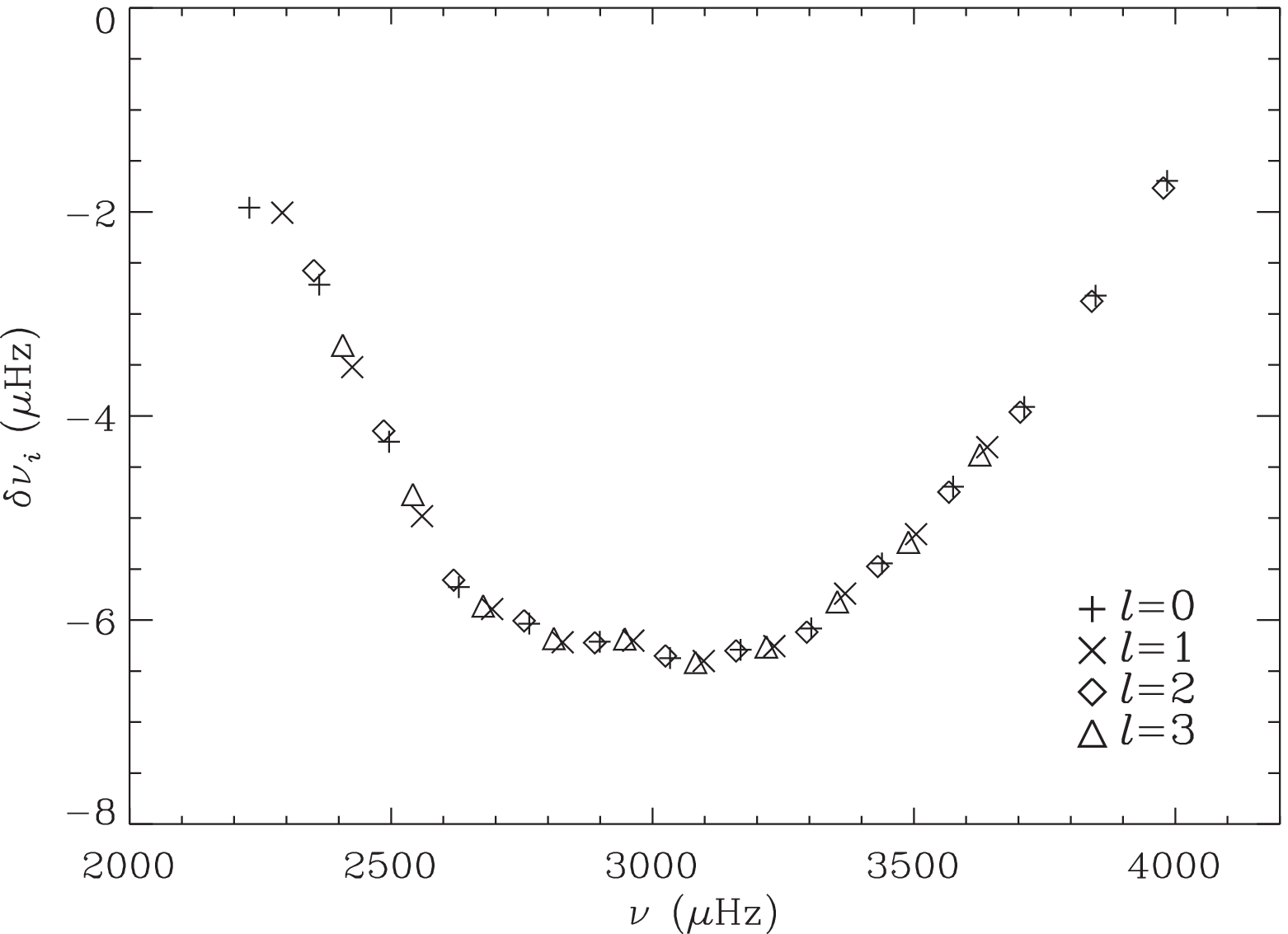}
\caption{
The symbols denote contributions $\delta\nu_{\boldsymbol i}$ to the frequencies
$\nu_{\boldsymbol i}$ produced by the acoustic glitches of the Sun
\citep[see also][]{hg09b}.}
\label{f:all_glitches}
\end{figure}

Following \citet{hg07b} we estimate the components of the \HL{displacement eigenfunction $\bxi$
of a mode of oscillation} of ${\cal M}_{\rm s}$, and \HL{the} divergence, in separated form as products of 
spherical harmonics and functions of radius $r$, using the
(hybrid) JWKB asymptotic approximation \citep[e.g.][]{g07} for 
high order $n$:
\begin{equation}
\xi\simeq\left(\frac{K}{r^2\rho}\right)^{1/2}\cos\psi\,,\quad
{\rm div}\bxi\simeq
     \left(\frac{\pi\omega^3|x|}{\gamma_1 pcr^2K}\right)^{1/2}{\rm Ai}(-x)\,,
\label{e:divxi}
\end{equation}
where $\xi(r)$ is the $r$-dependent factor in the vertical component of $\bxi$,
\HL{having  effective vertical wavenumber $K$}, and $\omega=2\pi\nu$ is the angular frequency of oscillation;
the argument $x$ of the Airy function Ai is given by
\begin{equation}
x:={\rm sgn}(\psi)\big\vert\frac{3}{2}\psi{\big\vert}^{2/3}
\end{equation}
in terms of the phase $\psi(\tau)=\int K\,{\rm d}r$, which we approximate
using a plane-parallel polytropic envelope of index $m$:
\begin{equation} 
\mbox{$\psi(\tau)\simeq$} \left\{
\begin{array}{lll} 
  \mbox{$\kappa\omega\tilde\tau-(m+1)\cos^{-1}\left(\frac{m+1}{\omega\tilde\tau}\right)$}&
  \mbox{for $\tilde\tau>\tau_{\rm t}$}\,,\\ \\ 
  \mbox{$|\kappa|\omega\tilde\tau-(m+1)\ln\left(\frac{m+1}{\omega\tilde\tau}+|\kappa|\right)$}&
  \mbox{for $\tilde\tau\le\tau_{\rm t}$}\,, 
\end{array} \right.
\label{eq:phase}
\end{equation}
in which 
$\tilde\tau=\tau+\omega^{-1}\epsilon$, with $\epsilon$ being a phase constant,
and $\tau_{\rm t}$ is the associated acoustical depth of the upper turning point, at which
the wavenumber $K$ vanishes.
The function
\begin{equation}
\kappa(\tau)=\left[1-\left(\frac{m+1}{\omega\tilde\tau}\right)^2\right]^{1/2}\,
\end{equation}
results from approximating $K$ as $c^{-1}(\omega^2-\omega^2_{\rm c})^{1/2}$
in which the acoustic cutoff frequency $\omega_{\rm c}$ is approximated by 
$(m+1)/\tilde\tau$.  \HL{Following \cite{hg07b} we take $m=3.5$.}
The Airy function must be adopted in the expression for div$\bxi$, which 
appears in the integral for $\delta_\gamma{\cal K}$ in equation~(\ref{e:divxi}),
because the upper turning point of the highest-frequency modes is within the 
He$\,$I ionization zone where $\delta\gamma_1$ is nonzero. It is adequate to 
use the sinusoidal (JWKB) expression for both $\xi$ and the horizontal
component of the displacement $\bxi$ -- which is determined as a horizontal
derivative in div$\,\bxi$ -- in computing the inertia, given by 
equation~(\ref{e:inertia}), because almost all of the integral comes from 
regions far from the turning points. It is approximated by 
${\cal I}\simeq\frac{1}{2}T\omega-\frac{1}{4}(m+1)\pi$ \citep{hg07b}, where
$T=\tau(0)$ 
is the acoustic radius of the star. The phase factor $\epsilon$ was
introduced to take some account of the variation with $\omega$ of the location of the upper
turning point.

Inserting these expressions into equations~(\ref{e:vp})--(\ref{e:K1}) \HL{yields the following 
approximation to} the helium-glitch frequency component: 
\begin{eqnarray}
\delta_\gamma\nu&=&-\sqrt{2\pi}A_{\rm II}\Delta^{-1}_{\rm II}
\left[\nu+\textstyle\frac{1}{2}(m+1)\nu_0\right]\cr
&&\hspace{-8pt}
\times\Bigl[\tilde\mu\tilde\beta\int_0^T\kappa^{-1}_{\rm I}
{\rm e}^{-(\tau-\tilde\eta\tau_{\rm II})^2/2\tilde\mu^2\Delta^2_{\rm II}}|x|^{\frac{1}{2}}
|{\rm Ai}(-x)|^2\,{\rm d}\tau\cr
&&\;\;+\int_0^T\kappa^{-1}_{\rm II}
{\rm e}^{-(\tau-\tau_{\rm II})^2/2\Delta^2_{\rm II}}|x|^{\frac{1}{2}}
|{\rm Ai}(-x)|^2\,{\rm d}\tau\Bigr]\,,
\label{eq:delgamnu}
\end{eqnarray}
where $\kappa_i:=\kappa(\tau_i)$, and where we have introduced a frequency
amplitude factor $A_{\rm II}=\frac{1}{2}\Gamma_{\rm II}T^{-1}$.

There are three additional components to $\Delta_2\nu_i$ that we must consider.
The first is due to the abrupt variation in the vicinity of the base of the 
convection zone at $\tau_{\rm c}$. We model it with a discontinuity in 
$\omega^2_{\rm c}$ at $\tau_{\rm c}$ coupled with an exponential relaxation
to the smooth model ${\cal M}_{\rm s}$ in the radiative zone beneath, with 
acoustical scale time $\tau_0 \HL{= 80{\rm s}}$, 
as did \citet{hg07b}. This leads to
\begin{eqnarray}
\delta_{\rm c}\nu&\simeq&A_{\rm c}\nu_0^3\nu^{-2}
   \left(1+1/16\pi^2\tau_0^2\nu^2\right)^{-1/2}\cr
&\times&\hspace{-3pt}\left\{\cos[2\psi_{\rm c}+\tan^{-1}(4\pi\tau_0\nu)]
      \!-\!(16\pi^2\tilde{\tau}_{\rm c}^2\nu^2\!+\!1)^{1/2}
\right\}\,,
\label{eq:delcnu}
\end{eqnarray}
where $\psi_{\rm c}:=\psi(\tau_{\rm c})$ and 
$\tilde\tau_{\rm c}:=\tilde\tau(\tau_{\rm c})$, and $A_{\rm c}$ is proportional
to the jump in $\omega^2_{\rm c}$.  

The other two components, whose sum we denote by $\delta_{\rm u}\nu_i$,  
contain a part that is generated in the very outer layers of the star --
by the ionization of hydrogen, the abrupt stratification of the upper
superadiabatic boundary layer of the convection zone, and by nonadiabatic
processes and Reynolds-stress perturbations associated with the oscillations,
which are difficult to model \citep[e.g.][]{rosen95, h10} -- and a part
that results from the incomplete removal of the smooth component when taking
a second difference. The latter \HL{was obtained} from equation~(\ref{e:asymp}),
\HL{and} is given approximately 
\HL{by the second derivative of 
$\nu_{{\rm s}\boldsymbol{i}}$ with respect to $n$, regarded as a continuous variable, 
retaining only the leading term.}
The degree-dependent term is much smaller than the other, and it is adequate 
here to regard the entire contribution as part of the essentially 
degree-independent upper (near-surface) glitch term, even though 
it actually arises in part from refraction in the radiative interior. 
We approximate it as a series in inverse powers of $\nu$, truncated at 
the cubic order:
\begin{equation}
\Delta_2\delta_{\rm u}\nu_{\boldsymbol i}=\sum_{k=0}^3a_k\nu_{\boldsymbol i}^{-k}\,.
\label{e:del2poly}
\end{equation}
We appreciate that in principle there should be an additional contribution
from the stellar atmosphere which, because it is produced far in the upper
evanescent region of the mode, is a high power of $\nu$ \citep{cdg80}.
However, for the Sun and Sun-like stars its contribution to the second
differences, used for determining $\Gamma_i$, is small, as can be adduced 
from the work by \citet{kbc08}. Its effect on the fitting of the smooth 
components $\nu_{{\rm s}{\boldsymbol i}}$ is \HL{mainly} to distort the values of $B$, $E$ and $I$. 
However, these coefficients are not used for the $t_\odot$ and $Y_0(Z_0)$ 
calibration. Accordingly, we can safely ignore this surface contribution.
\citet{dbc11} have recently illustrated this general point \HL{with specific numerical examples}. 

The complete \HL{expression for the} second difference
\begin{equation}
\Delta_{2{\boldsymbol i}}\nu
\simeq\Delta_{2}(\delta_\gamma\nu_{\boldsymbol i}+\delta_{\rm c}\nu_{\boldsymbol i}+\delta_{\rm u}\nu_{\boldsymbol i})
    =:g_{\boldsymbol i}(\nu_{\boldsymbol j}; \eta_\alpha)
\label{e:all_glitches}
\end{equation}
was then fitted to the second differences of the solar, or solar-model, 
frequencies to determine the coefficients
\HL{
$\eta_\alpha=(A_{\rm II}$, $\Delta_{\rm II}$, $\tau_{\rm II}$, 
$\epsilon_{\rm II}$, $A_{\rm c}$, $\tau_{\rm c}$, $\epsilon_{\rm c}, a_0, a_1, a_2, a_3),\, 
\alpha=1,...,11$. 
}
From the outcome, putative frequency contributions
$\delta_{\rm u}\nu_i$ were obtained by summing the 
second differences~(\ref{e:del2poly}) to yield
\begin{eqnarray}
\delta_{{\rm u}}\nu_{\boldsymbol i}
&\simeq&\tilde A+\tilde B\nu_i\cr
&\!\!+\!\!&{\frac{1}{2}}\left[a_0\nu_i^2+2a_1\nu_i(\ln\nu_{\boldsymbol i}\!-\!1)-2a_2\ln\nu_{\boldsymbol i}+a_3\nu_i^{-1}\right]\cr
& &\times(\nu_{n+1,0}-\nu_{n-1,0})\cr
&\equiv&\tilde A+\tilde B\nu_{\boldsymbol i}+F_{{\rm u}{\boldsymbol i}}\,.
\label{e:delnu_surf}
\end{eqnarray}
The initially arbitrary constants of summation $\tilde A$ and $\tilde B$ were
selected in such a way as to minimize the $L_2$ norm of $\delta_{\rm u}\nu_{\boldsymbol i}$,
namely $\sum_{\boldsymbol i}(\tilde A+\tilde B\nu_{\boldsymbol i}+F_{{\rm u}{\boldsymbol i}})^2$, 
as did \citet{hg09a}.

The fitting of the second-differences was accomplished by minimizing
\begin{equation}
E_{\rm g}=(\Delta_{2{\boldsymbol i}}\nu-g_{\boldsymbol i}){\rm C}^{-1}_{\Delta {\boldsymbol {ij}}}(\Delta_{2{\boldsymbol j}}\nu-g_{\boldsymbol j})
\label{e:minsecdiff}
\end{equation}
using the value of $\nu_0$ obtained from the fitting of expression\,(\ref{e:asymp}).
(That fitting was accomplished by minimizing the appropriately weighted
mean-square difference $E_{\rm s}$  \HL{-- defined in \S\,2.4 -- } from the smooth 
frequencies $\nu_{{\rm s}{\boldsymbol i}}$,
which are themselves derived from the raw frequencies by subtracting the glitch
contribution obtained by minimizing $E_{\rm g}$; the two minimizations were
carried out iteratively in tandem.)
Here C$^{-1}_{\Delta {\boldsymbol {ij}}}$ is the $({\boldsymbol i},{\boldsymbol j})$ 
element of the inverse of the 
covariance matrix C$_\Delta$ of the observational errors in 
$\Delta_{2{\boldsymbol i}}\nu$, computed, perforce, under the assumption that the 
errors in the frequency data $\nu_{\boldsymbol i}$ are independent. 
The resulting covariance matrix C$_{\eta\alpha\gamma}$ of the errors 
in $\eta_\alpha$ was established by Monte Carlo simulation, using 6000
realizations of Gaussian-distributed errors in the raw data with variance in
accord with the published standard errors. In carrying out the simulations
we omitted the surface term $\delta_{\rm u}\nu_{\boldsymbol i}$, which has
insignificant influence on the statistics.


\begin{figure}
\centering
\includegraphics[scale=0.75]{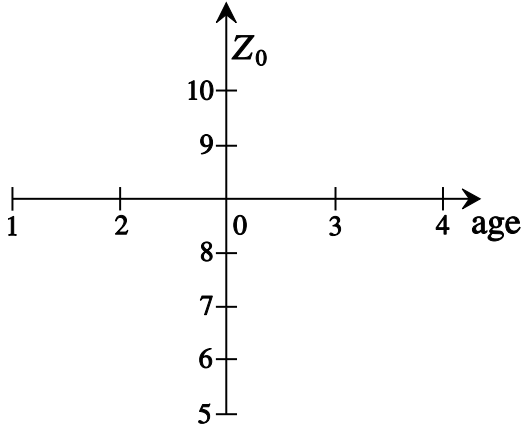}
\caption{
Denotation of the eleven solar models which we have used for calculating 
the partial derivatives $H_{\alpha\beta}$, calibrated to a present radius 
${\rm R}_\odot=6.9599\times10^{10}$ cm and luminosity 
${\rm L}_\odot=3.846\times10^{33}$erg$\,$s$^{-1}$.  The `central model' 
is Model~0; the sequence of the five models (0-4) has a constant value of $Z_0=0.02$ 
but varying age $t_\star$ (4.15, 4.37, 4.60, 4.84, 5.10)$\,$Gy; the second sequence, of seven models  
(0,5-10), has constant age $t_\star=4.60\,$Gy but varying $Z_0$ 
(from 0.016 to 0.022 in \HL{uniform} steps of 0.001).
}
\label{f:model_grid}
\end{figure}


The outcome of the fitting to the BiSON data
is displayed in Fig.\,\ref{f:BiSON}: 
the upper panel displays the second differences, 
together with the complete fitted formula~(\ref{e:all_glitches}) (solid curve) and
its individual smooth frequency contribution $\delta_{\rm u}\nu_i$ 
\HL{estimated by equation~(\ref{e:del2poly})};
the corresponding oscillatory frequency contributions 
\HL{$\Delta_2\delta_\gamma\nu$}
(dotted and solid curves for the two stages of helium ionization) and
\HL{$\Delta_2\delta_{\rm c}\nu$}
(dot-dashed curve) are illustrated in the lower panels 
of Fig.\,\ref{f:BiSON}.
All the frequencies displayed in the figure have been used in equation
(\ref{e:minsecdiff}) for fitting expression~(\ref{e:all_glitches}).

In Fig.\,\ref{f:all_glitches} is displayed the sum of all acoustic 
\HL{glitch contributions to the frequencies estimated 
by fitting equation~(\ref{e:all_glitches}) to the low-degree solar frequencies observed by
BiSON \citep{basu07}. 
}


\begin{figure}
\centering
\includegraphics[width=1.00\linewidth]{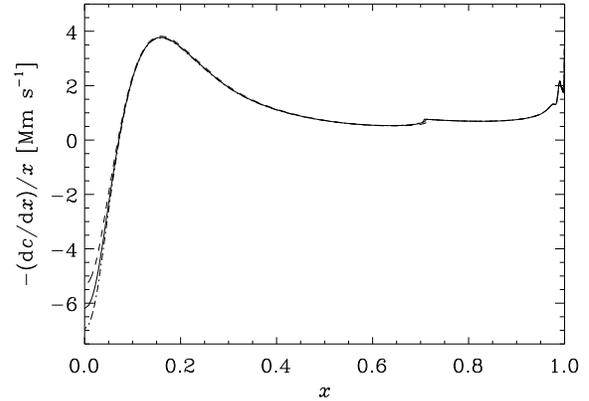}
\caption{
Integrand $-({\rm d}c/{\rm d}x)/x$ of $A$ as a function of radius fraction $x:=r/R$
for the calibrated solar models 5, 0 and 10 with varying $Z_0$ at constant age 
$t_\odot=4.60\,$Gy. The dashed, solid and dot-dashed curves are the results
for models 5 ($Z_0$=0.016), 0 ($Z_0=0.020$) and 10 ($Z_0$=0.022) respectively.
\HL{Note that the sensitivity to $Z_0$ lies mainly near the centre; the same 
is so for the sensitivity to $t_\star$ \citep{gn90}}.
} 
\label{f:AZ0}
\end{figure}


   \subsection{Calibration for age and chemical composition}
   \label{s:calibration}
We subtract the glitch contributions $\delta\nu_{\boldsymbol i}$ from the 
full frequencies to obtain corresponding glitch-free frequencies 
$\nu_{{\rm s}{\boldsymbol i}}$.  The procedure
is carried out for the solar observations, for the eigenfrequencies of the 
reference solar model, and for the grid of models used for evaluating derivatives
of the fitting parameters with respect to $t_\star$ and \HL{$Z_0$  
(see Fig.\,\ref{f:model_grid})}. 
Then we iterate the parameters defining the reference model
by minimizing 
$E_{\rm s}:=(\nu_{{\rm s}{\boldsymbol i}}-S_{\boldsymbol i}){\rm C}^{-1}_{{\rm s}{\boldsymbol {ij}}}(\nu_{{\rm s}{\boldsymbol j}}-S_{\boldsymbol j})$,
where C$_{\rm s}$ is the covariance matrix of the 
statistical errors in $\nu_{{\rm s}{\boldsymbol i}}$, which are determined from the
independent observational errors in $\nu_{\boldsymbol i}$ and the covariance
matrix C$_{\eta_{\alpha\beta}}$, to obtain both the coefficients $\xi_\beta$
and the covariance matrix C$_{\xi\beta\delta}$ of their errors.
In this iteration process for $\xi_\beta$, only glitch-free frequencies 
$\nu_{{\rm s}{\boldsymbol i}}$ with $k=n+\frac{1}{2}l\ge15$ were considered, because 
the asymptotic expression~(\ref{e:asymp}) is not sufficiently accurate for lower
values of $k$.
Each component of $\xi_\beta$ is an integral of a function of the equilibrium 
stratification. Some of these are displayed in Fig.\,\ref{f:integrals}.
The integrals $A, C$ and $F$ are those of particular importance to our 
analysis, because $C$ and $F$ are dominated by conditions in the core, 
and, although the contributions to $A$ from the core and the rest of the 
star are roughly equal in magnitude (and potentially have opposite signs), 
the contribution from the envelope is relatively insensitive to $t_\star$ 
\citep{gn90} and $Z_0$ (Fig.\,\ref{f:AZ0}). 
The integrands in the remaining integrals are either more evenly 
distributed throughout the Sun or are concentrated near the surface.

\begin{figure*}
\centering
\includegraphics[width=0.40\linewidth,angle=90]{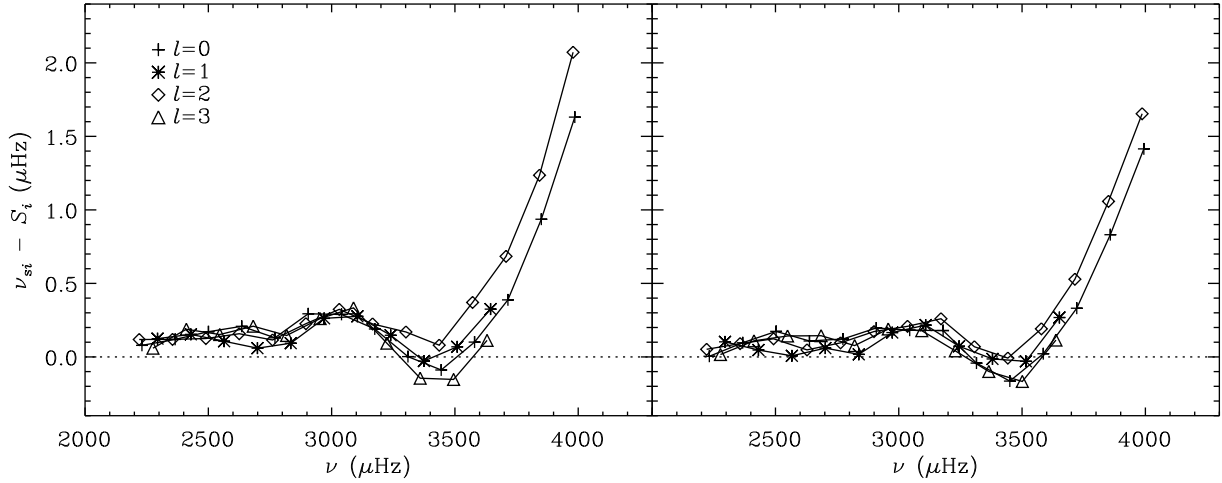}
\caption{Differences between smoothed frequencies $\nu_{{\rm s}{\boldsymbol i}}$,  
of the Sun (left) and central Model 0 (right), and the correspondingly fitted asymptotic
expression $S_{\boldsymbol i}$ given by~(\ref{e:asymp}). Modes of like degree $l$ 
are connected by solid lines.  \HL{Contrary to the opinions of some commentators, 
the curves are not like the oscillatory contribution from $\delta\gamma_1/\gamma_1$, for rather than decreasing, the amplitude of 
the undulation increases  with increasing $\nu$  (cf. Fig. 2), indicating discrepancies in 
the very outer layers.}
} 
\label{f:nus_residuals}
\end{figure*}

The differences between the 
smoothed frequencies $\nu_{{\rm s}{\boldsymbol i}}$ and the fitted asymptotic 
expression $S_{\boldsymbol i}$ given by equation~(\ref{e:asymp}) 
are displayed in Fig.~\ref{f:nus_residuals} for the BiSON data
(left panel) and for the central model m0 (right panel).


\begin{figure*}
\centering
\includegraphics[width=0.51\linewidth,angle=90]{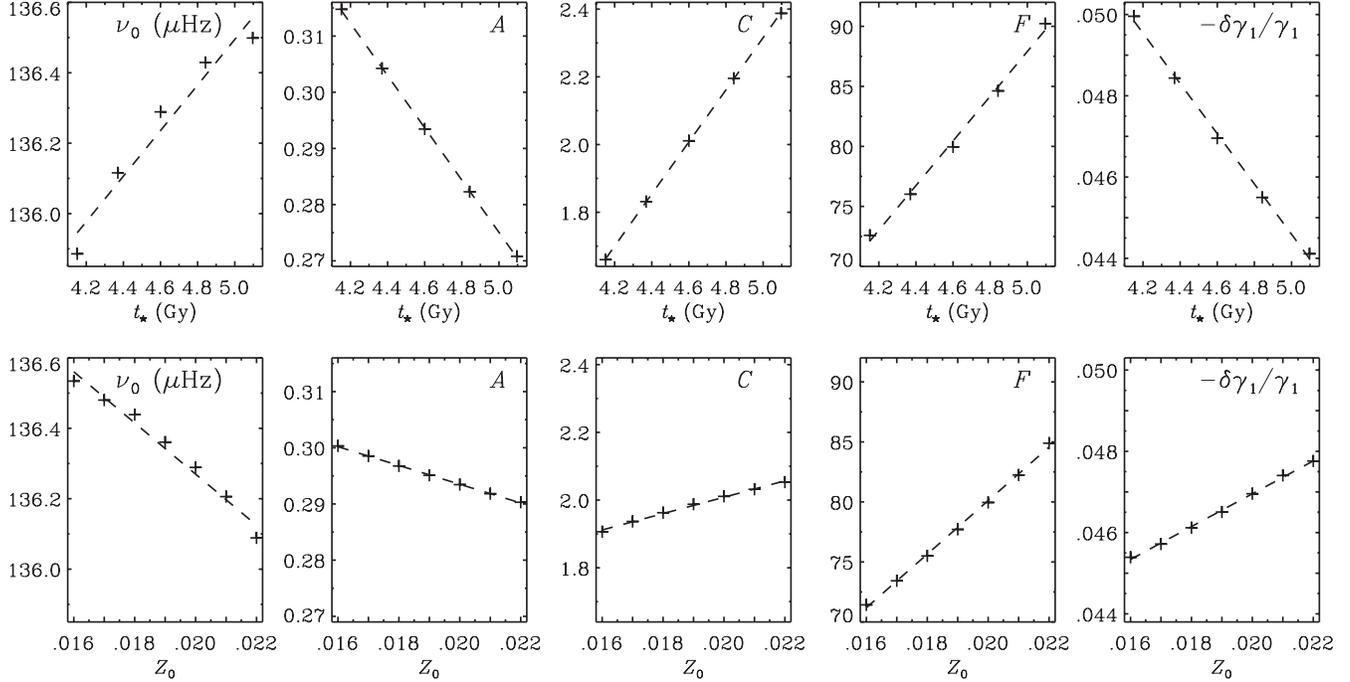}
\caption{
Seismic parameters $\nu_0, A, C, F$ and $-\delta\gamma_1/\gamma_1$ of eleven calibrated 
solar models from which their partial derivatives $H_{\alpha j}$ 
with respect to stellar age $t_\star$ and initial heavy-element 
abundance $Z_0$, listed in Table~\ref{t:hij}, are obtained.
} 
\label{f:calib-models}
\end{figure*}

We have carried out age calibrations using various combinations of the parameters 
\begin{equation}
\zeta_\alpha=(\hat A,\hat C,\hat F,-\delta\gamma_1/\gamma_1),\qquad \alpha=1,...,4\,,  
\label{eq:xi}
\end{equation}
where $-\delta\gamma_1/\gamma_1=A_{\rm II}/\sqrt{2\pi}\nu_0\Delta_{\rm II}$ is
a measure of the maximum depression in $\gamma_1$ in the second helium 
ionization zone, and which for convenience we sometimes denote by $\hat\Gamma$.
The values of  $-\delta\gamma_1/\gamma_1$ and the asymptotic coefficients $A, C, F$ 
appearing in expression~(\ref{e:asymp}), \HL{determined 
from the observed seismic frequencies}, 
are listed in Table~\ref{t:BiSON_coefficients} for the Sun, and are plotted in 
Fig.~\ref{f:calib-models} for the eleven calibrated grid models. 
Presuming, as is normal, that the reference model is parametrically close 
to the Sun, we first carry out a single iteration by approximating the 
reference value $\zeta^{\rm r}_\alpha$ by a two-term Taylor 
expansion about the value $\zeta^\odot_\alpha$ of
the Sun:
\begin{equation}
\zeta^{\rm r}_\alpha=\zeta^{\odot}_\alpha
   -\left(\frac{\partial\zeta_\alpha}{\partial t_\star}\right)_{\!\!Z_0}\Delta\,t_\star
   -\left(\frac{\partial\zeta_\alpha}{\partial Z_0}\right)_{\!\!t_\star}\Delta Z_0
   +\epsilon_{\zeta\alpha}\, ,
\end{equation}
where $\Delta\,t_\star$ and $\Delta Z_0$ are the deviations of the age $t_\odot$ 
and initial heavy-element abundance $Z_0$ of the Sun from the corresponding 
values of the reference model; $\epsilon_{\zeta\alpha}$
are the formal errors in the calibration parameters, whose covariance
matrix C$_{\zeta\alpha\beta}$ can be derived from C$_{\xi\beta\delta}$
and C$_{\eta\alpha\gamma}$. A (parametrically local) maximum-likelihood 
fit then leads to the following set of linear equations: 
\begin{equation}
H_{\alpha j}{\rm C}^{-1}_{\zeta\alpha\beta}H_{\beta k}\Theta_{0k}=
H_{\alpha j}{\rm C}^{-1}_{\zeta\alpha\beta}\Delta_{0\beta}\,,
\label{eq:calib1}
\end{equation}
in which $\Theta_k=(\Delta t_\star, \Delta Z)+\epsilon_{\Theta k}=
\Theta_{0k}+\epsilon_{\Theta k}$, $k=1,2$, is the solution vector subject 
to (correlated) errors $\epsilon_{\Theta k}$, and
$\Delta_\beta=\zeta^\star_\beta-\zeta^{\rm r}_\beta+\epsilon_{\zeta\beta}
=\Delta_{0\beta}+\epsilon_{\zeta\beta}$; the partial derivatives are denoted by
$H_{\alpha j}=[(\partial\zeta_\alpha/\partial t_\star)_{Z_0},
(\partial\zeta_\alpha/\partial Z)_{t_\star}]$, $j=1,2$.

A similar set of equations is obtained for the formal errors 
$\epsilon_{\Theta k}$: 
\begin{equation}
H_{\alpha j}{\rm C}^{-1}_{\zeta\alpha\beta}H_{\beta k}\epsilon_{\Theta k}=
H_{\alpha j}{\rm C}^{-1}_{\zeta\alpha\beta}\epsilon_{\zeta\beta}\,,
\label{eq:calib2}
\end{equation}
from whose solution the error covariance matrix 
C$_{\Theta kq}=\overline{\epsilon_{\Theta k}\epsilon_{\Theta q}}$ can be 
computed.

\begin{table}
\centering
\caption{Asymptotic coefficients for the Sun obtained from (linear) fitting 
to the glitch-free (smoothed) BiSON frequencies $\nu_{\rm s}$
the expression~(\ref{e:asymp}) by 
\HL{weighted}
least squares $E_{\rm s}$.
}
\begin{tabular}{ccccc}
\noalign{\smallskip}
\noalign{\smallskip}
\hline
\noalign{\smallskip}
 \hfil $\nu_0$ ($\mu$Hz)\hfil&
 \hfil $A$\hfil&
 \hfil $C$\hfil&
 \hfil $F$\hfil&
 \hfil $-\delta\gamma_1/\gamma_1$\hfil\\
\hline
  136.71&0.3005&1.912&69.83&0.04538\\
\noalign{\smallskip}
\hline
\end{tabular}
\label{t:BiSON_coefficients}
\end{table}

The partial derivatives $H_{\alpha j}$ were obtained from the set of 
eleven calibrated evolutionary models (see Fig.\,\ref{f:model_grid})
of the Sun that were used in a similar calibration by \cite{hg07a}. 
The models
were computed with the evolutionary programme by \cite{jcd08}, 
adopting the Livermore equation of state and the OPAL92 opacities. 
The set comprises two sequences:
one has a constant value of the heavy-element abundance 
$Z_0=0.020$ but varying age 
($t_\star=4.15,...,5.10\,$Gy 
\HL{in uniform steps of 0.05 in}
$\ln t_\star$); 
the other has constant age $t_\star=4.60\,$Gy
but varying $Z_0$ ($Z_0$=0.016,...,0.022 
\HL{in uniform steps of 0.001}). 
Note that, for prescribed relative abundances of heavy elements,
the condition that the luminosity and radius of the Sun agree with
observation defines a functional relation between $Y_0, Z_0$ and \HL{$t_\star$ 
amongst the models}.
In Fig.\,\ref{f:calib-models} are plotted the seismic parameters $A, C, F$ 
and $-\delta\gamma_1/\gamma_1$ of the eleven models, each calibrated to 
the solar radius 
and luminosity, for determining the partial derivatives 
\HL{$H_{\alpha j}$}
of $A, C, F$ and $-\delta\gamma_1/\gamma_1$ with respect to stellar age $t_\star$ and 
\HL{initial}
heavy-element abundance $Z_0$.
The values of the partial (logarithmic) derivatives $H_{\alpha j}$ so obtained are listed 
in Table\,\ref{t:hij}.
Notice that within the range of model parameters that we have considered, the
derivatives are almost constant.

\begin{table*}
\centering
\caption{Partial logarithmic derivatives ($H_{\alpha j}$) obtained 
from the two sets of calibrated evolutionary models for the Sun denotated in Fig. 4. They 
were determined by linear regression about the central Model~0.  
}
\begin{tabular}{cccccccccc}
\noalign{\smallskip}
\noalign{\smallskip}
\hline
 \hfil$\left(\frac{\partial\ln\nu_0}{\partial\ln t_\star}\right)_{Z_0}$\hfil&\hfil$\left(\frac{\partial\ln\nu_0}{\partial\ln Z_0}\right)_{t_\star}$\hfil&
 \hfil$\left(\frac{\partial\ln A}{\partial\ln t_\star}\right)_{Z_0}$\hfil&\hfil$\left(\frac{\partial\ln A}{\partial\ln Z_0}\right)_{t_\star}$\hfil&
 \hfil$\left(\frac{\partial\ln C}{\partial\ln t_\star}\right)_{Z_0}$\hfil&\hfil$\left(\frac{\partial\ln C}{\partial\ln Z_0}\right)_{t_\star}$\hfil\\
\hline
\,0.0220\,&\,-0.00997\,&\,-0.733\,&\,-0.107\,&\, 1.771\,&\,0.231\,\\
\hline
\hline
 \hfil$\left(\frac{\partial\ln F}{\partial\ln t_\star}\right)_{Z_0}$\hfil&\hfil$\left(\frac{\partial\ln F}{\partial\ln Z_0}\right)_{t_\star}$\hfil&
 \hfil$\left(\frac{\partial\ln(-\delta\gamma_1/\gamma_1)}{\partial\ln t_\star}\right)_{Z_0}$\hfil&
 \hfil$\left(\frac{\partial\ln(-\delta\gamma_1/\gamma_1)}{\partial\ln Z_0}\right)_{t_\star}$\hfil&
 \hfil$\left(\frac{\partial\ln Y_0}{\partial\ln t_\star}\right)_{Z_0}$\hfil&\hfil$\left(\frac{\partial\ln Y_0}{\partial\ln Z_0}\right)_{t_\star}$\hfil\\
\hline
\,1.057 \,&\,0.539   \,&\,-0.607\,&\,0.163 \,&\,-0.173\,&\,0.334\,\\
\hline
\end{tabular}
\label{t:hij}
\end{table*}

   \section{Results}
   \label{sec:results}
Provided that the reference model is close to the Sun, the single iteration
described in the previous section should provide as reliable an estimate
of ($\Delta t_\odot,\;\Delta Z_0$) as the calibration is currently able to provide.
We therefore discuss at first the results of single iterations.
Calibrations were carried out using different combinations of the parameters 
$\zeta_\alpha$ and two different reference models. They are summarized in 
Table\,\ref{t:calibration}.
The older reference model is the central `Model~0' 
which has age $t_\star=4.600\,$Gy; the second is `Model~2', which has an age
$t_\star=4.370\,$Gy. 
\HL{Because the acoustic properties of the stars in the grid vary almost linearly with 
$t_\star$ and $Z_0$ the error covariance matrices associated with the single iteration 
are indistinguishable.}  
We adopted the same physics as in Model S 
\citep{jcd96} in the
evolutionary calculations 
\HL{of}
both models. We notice 
by comparing rows 4 and 6 with rows 5 and 7
in Table\,\ref{t:calibration} that 
calibrations without $\delta\gamma_1/\gamma_1$ are less 
stable to a change in the reference model than are the calibrations including 
$\delta\gamma_1/\gamma_1$. 
\HL{They are possibly}
less reliable, for the reasons explained in the introduction, 
although the result may perhaps be simply a symptom of slower convergence.

\begin{figure*}
\centering
\includegraphics[angle=90,width=0.65\linewidth]{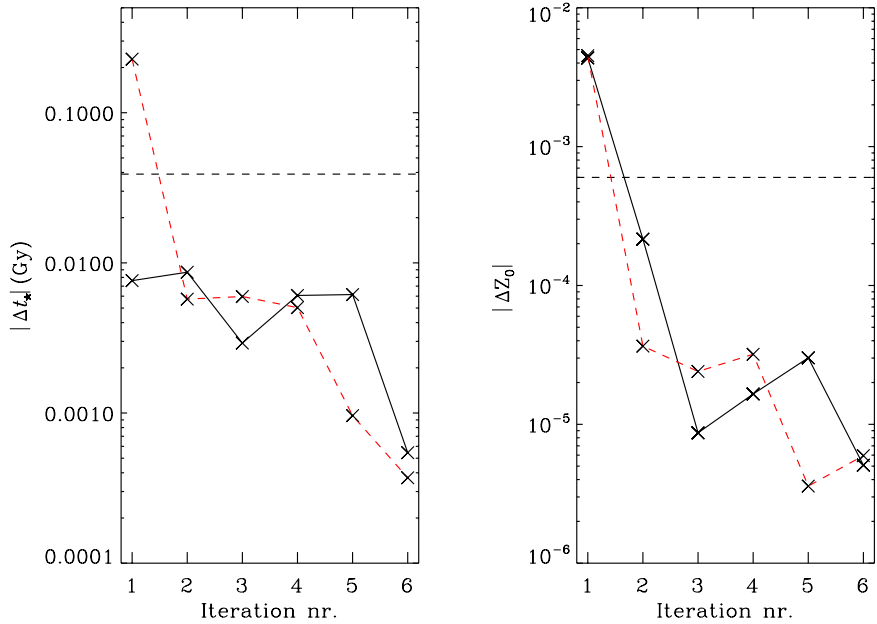}
\caption{
Corrections of the solar age $t_\odot$ 
\HL{(left panel)}
and initial heavy-element abundance $Z_0$
\HL{(right panel)}
for six calibration iterations and the combination $\hat A, \hat C, \hat F$ and 
$-\delta\gamma/\gamma$ for two reference models with an initial
age of $t=4.60\,$ (solid black curves) and $t=4.37\,$Gy (dashed red curves) and
initial heavy-element abundance $Z_0=0.02$. The horizontal dashed lines are the
estimated 1$\,\sigma$ error bars of the calibrated age and initial heavy-element 
abundance.
} 
\label{f:caliter}
\end{figure*}

\begin{figure}
\centering
\includegraphics[width=1.00\linewidth]{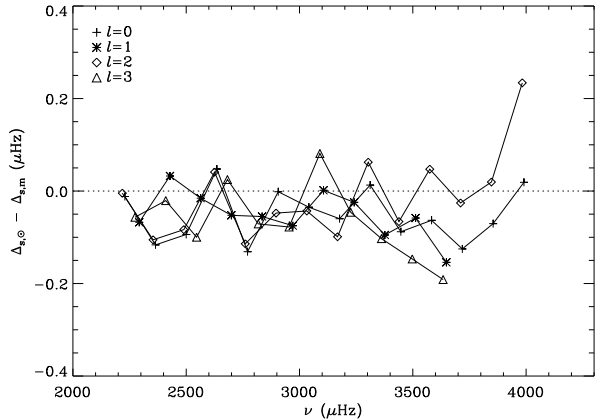}
\caption{
\HL{Differences between residuals $\Delta_{\rm s}:=\nu_{{\rm s}\boldsymbol i}-S_{\boldsymbol i}$ 
\HL{obtained from the frequencies of} the Sun, $\Delta_{{\rm s},\odot}$, and the }
\HL{calibrated  model after five iterations from}\HL{  reference Model~0, 
$\Delta_{{\rm s, m}}$, where expression $S_{\boldsymbol i}$ given 
by Eq.\,(\ref{e:asymp}) was fitted to the smoothed frequencies $\nu_{{\rm s}\boldsymbol i}$.  }
} 
\label{f:diff_residuals}
\end{figure}

To ascertain whether the
entire calibration procedure converges, we have performed several 
additional iterations. At each iteration 
the corrections $\Delta t_\star$ and $\Delta Z_0$ are used to define parameters
of a new reference model, which is then constructed by
performing another evolutionary calculation, followed by the evaluation of a new set
of corrections $\Delta t_\star$ and $\Delta Z_0$ as before. We repeated this 
for five iterations, for each of the two reference models, obtaining
the two `final' reference models, listed in Table~\ref{t:results}, 
for two different combinations of $\zeta_\alpha$, \HL{whose present 
heavy-element abundance $Z$ is}  displayed in Fig.\,\ref{f:Zprofile}. 
The \HL{progressive} corrections $\Delta t_\star$ and $\Delta Z_0$ are plotted in 
Fig.\,\ref{f:caliter}.
In carrying out the iterations we did not recompute the partial derivatives
$H_{\alpha j}$ and the corresponding error covariance matrices. To have done so
would have been computationally much more expensive, would have been likely not to have 
speeded up convergence by very much, 
\HL{and would not have altered the final solution.  The final residuals 
$\Delta_{\rm s}:=\nu_{{\rm s}{\boldsymbol i}}-S_{\boldsymbol i}$ are barely distinguishable 
from those from the Sun, as illustrated in Fig.\,\ref{f:diff_residuals}.}

\begin{table*}
\centering
\caption{Age calibrations with different combinations of $\zeta_\alpha$ and
for the two reference models: Model\,0 with an age $t_\star=4.60\,$Gy 
and Model\,2 with an age $t_\star=4.37\,$Gy, both computed with an
initial heavy-element abundance $Z_0\,=\,0.02$ adopting the \citet{greno93} 
solar composition. The first three columns show the
results adopting Model~0 as the reference model, the fourth, fifth and sixth
columns display the results for Model~2. \HL{The remaining three columns are
the values of the error covariance matrix C$_\Theta$ for both reference models.}
}
\begin{tabular}{lccccccccc}
\noalign{\smallskip}
\noalign{\smallskip}
\hline
 \hfil$\zeta_\alpha$\hfil&
 \hfil $t_\odot$ (Gy)\hfil&
 \hfil$Z_0$\hfil&
 \hfil$Y_0$\hfil&
 \hfil $t_\odot$ (Gy)\hfil&
 \hfil$Z_0$\hfil&
 \hfil$Y_0$\hfil&
 \hfil C$^{1/2}_{\Theta 11}$\hfil&
 \hfil$-(-{\rm C}_{\Theta 12})^{1/2}$\hfil&
 \hfil C$^{1/2}_{\Theta 22}$\hfil\\
\hline
$\hat A,\hat C,\hat F,-\delta\gamma_1/\gamma_1$&4.592&0.0156&0.252&4.597&0.0155&0.251&0.039&0.0013&0.0005\\
$\hat A,\hat F,-\delta\gamma_1/\gamma_1$       &4.580&0.0157&0.252&4.582&0.0156&0.251&0.045&0.0016&0.0006\\
$\hat C,\hat F,-\delta\gamma_1/\gamma_1$       &4.591&0.0157&0.252&4.595&0.0155&0.251&0.044&0.0004&0.0005\\
$\hat A,\hat C,-\delta\gamma_1/\gamma_1$       &4.597&0.0160&0.254&4.603&0.0160&0.253&0.045&0.0036&0.0008\\
$\hat A,\hat C,\hat F                  $       &4.619&0.0153&0.252&4.632&0.0151&0.248&0.095&0.0104&0.0013\\
$\hat A,\hat C                         $       &4.638&0.0147&0.246&4.654&0.0143&0.245&1.049&0.1791&0.0306\\
$\hat A,-\delta\gamma_1/\gamma_1       $       &4.588&0.0159&0.253&4.592&0.0158&0.253&0.149&0.0222&0.0039\\
\noalign{\smallskip}
\hline
\end{tabular}
\label{t:calibration}
\end{table*}


Error contours corresponding to the calibration from Model~0 in the first 
row of Table\,\ref{t:results} are plotted in Fig.\,\ref{f:errellipse}. Corresponding 
contours for Model~2 are the same, except that their centres are displaced 
to (4.603\,Gy, 0.0155).
One can adduce from our description of the analysis in \S\,\ref{s:calibration}
that our current treatment of the errors is not completely unbiassed, because, aside from 
$\nu_0$,  we assess the 
error covariances of the parameters defining the smooth and the glitch components 
independently; however, the potential bias is of the order of only $|\delta\nu_i/\nu_i|$ or less, 
which is small.

Fig.\,\ref{f:Zprofile} depicts the heavy-element profiles after five 
iterations from the two reference models 
\HL{(Models~0 and 2)}.
Both models have a surface value $Z_{\rm s}=0.0142\pm0.0005$, which is 
about 6\% higher than the value of $Z_{\rm s}=$0.0134 reported by \citet{asp09} and 
about 9\% smaller than the value of $Z_{\rm s}=0.0156\pm0.0011$ reported by 
\citet{caf09}.
The error-bars of Caffau's $Z_{\rm s}$ value, obtained from numerical simulations, 
is indicated by the 
\HL{shaded}
region.

The calibrated age inferred from Model~0 after five iterations 
is 4.604$\pm0.039\,$Gy, and that 
from Model~2 is 4.603$\pm0.039\,$Gy, using 
the parameter combination $\hat A, \hat C, \hat F$ and $-\delta\gamma_1/\gamma_1$.
The corresponding calibrations from \HL{M}odels 0 and 2 for the combination 
$\hat C, \hat F$ and $-\delta\gamma_1/\gamma_1$ 
are $4.602\pm0.044\,$Gy and $4.601\pm0.044\,$Gy, respectively.
Table\,\ref{t:results} summarizes the calibrations after five iterations from
reference 
\HL{Models~0 and 2}.


   \section{Discussion}
   \label{sec:discussion}
In attempting to estimate the main-sequence age of the Sun it is prudent to
adopt diagnostic quantities that are insensitive to properties that one
believes not to be directly pertinent. As the Sun evolves on the main sequence
it converts hydrogen into helium in the core. \HL{According to theoretical models it  
liberates energy at a rate whose dependence on time, measured in units of the age 
$t_\star$, }
is not very sensitive to uncertain parameters defining those models, such as the initial
heavy-element abundance $Z_0$, provided that the models have been calibrated
to reproduce the luminosity and radius observed today \HL{(cf. Gough, 1990b)}. The same is true of the
quantity of hydrogen consumed, mainly because the nuclear relations are dominated
by a single branch of the pp chain, namely ppI, for which there is a tight
link between fuel consumption and 
\HL{thermal}
energy release.
Therefore \HL{one would expect there to be} a robust link between main-sequence age and 
the total amount of hydrogen consumed: the integral
$\Delta H:=\int h(r)\rho r^2\,{\rm d}r:=4\pi\int\left[X_0-X(r)\right]\rho r^2\,{\rm d}r$
\HL{should be} a good indicator of the age $t_\odot$. It can be calibrated using seismic
diagnoses of the mean molecular mass $\mu(r)$ provided that processes other than
nuclear reactions that can change $X(r)$, such as gravitational settling and diffusion,
are taken adequately into account.

In perhaps its simplest form, solar evolution involves computing models of constant mass
in hydrostatic equilibrium. The models usually depend on three initial parameters: the 
\HL{initial}
abundances of, say, helium, $Y_0$, and the heavy elements, $Z_0$, and a 
mixing-length parameter $\alpha_{\rm c}$ which is normally held constant. It is usual to 
fix the relative abundances of all elements other than hydrogen and helium, a procedure 
which we too have adopted here. Demanding that the
luminosity and radius of the model agree with present-day observation relates two of
those parameters, say $\alpha_{\rm c}$ and $Y_0$, to the third, $Z_0$, for any $t_\star$. 
Thus one obtains a two-parameter set of potentially acceptable models, which here we 
characterize by the values of $t_\star$ and $Z_0$, and which we attempt to calibrate 
with helioseismic data.

Several diagnostics have been used in the past. As mentioned in the introduction, the 
first to be used for a full calibration were the two mean small separations $d_0$ and $d_1$
\citep{dog01}, averages over $n$ of $d_{n,0}$ and $d_{n,1}$, the hope being that the
differences in the way in which the two quantities sample the core would be adequate to
disentangle $t_\odot$ and $Z_0$. Unfortunately, given the precision of the data at the time,
that could not be accomplished to a useful precision. Moreover, by inspecting the 
dependence of the calibration on the range of frequencies over which the averages $d_0$ and
$d_1$ were determined, there was evidence of contamination by an oscillatory component
to the signatures from seismically abrupt variations of the stratification in and at the
base of the convection zone. This component is particularly visible in second- and
higher-order frequency differences with respect to $n$ \citep[e.g.][]{g90a, btc04, bm04}. 

There were several obvious improvements 
\HL{ to the original calibration based solely on fitting to  raw 
frequencies a smooth-asymptotic formula, such as equation~(\ref{e:asymp}) 
or derivatives of it, that were required to be put into place}  in order to obtain a more
reliable calibration. The first that we have made is to isolate much of the
signal from the abrupt variation \HL{ of the thermodynamic properties} in the convection zone. 
The intention was two-fold: 
First, by removing the oscillatory component from the stratification one is left
with a smooth model for which the simple asymptotic expression\,(\ref{e:asymp}) is more
nearly valid; second, its amplitude provides an independent measure of the helium
abundance $Y_{\rm s}$ in the convection zone \citep{hg07b} through the magnitude of the
depression in $\gamma_1$ in the ionization zones. The latter provides, via stellar-evolution
theory, the value of $Y_0$ -- and therefore $X_0(Z_0)$ -- in the core, which is required for
determining the hydrogen deficit $h(r)$. In carrying out the analysis, the variation in 
$\gamma_1$ has been represented by two Gaussian functions of acoustic depth, as 
\HL{recommended}
by \citet{dog02} and \citet{hg04}, which has been found to reproduce the oscillation frequencies 
more faithfully than either the simple discontinuity that was adopted by 
\citet{b04}, \citet{bm04} and \citet{mm10}, and the triangular 
form adopted by \citet{mt98, mt05} 
\HL{and \citet{vce06}}; 
presumed discontinuities in $\gamma_1$ or its derivatives cause the amplitude
of the predicted oscillatory feature to decay too slowly with frequency \citep{hg04}, which, 
although apparently not very deleterious for the Sun, may be a serious deficiency for
other stars.

\begin{figure}
\centering
\includegraphics[width=0.95\linewidth]{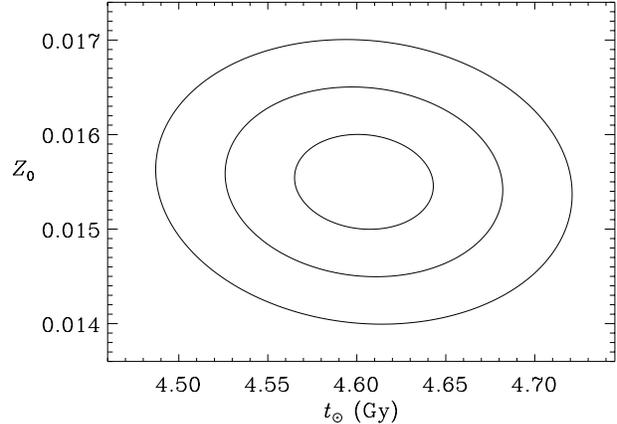}
\caption{
   Error ellipses for the calibration using the combination
   $\hat A, \hat C, \hat F$, $-\delta\gamma_1/\gamma_1$ and Model~0 
   as the reference model:
   solutions $(t_\odot,Z_0)$ satisfying the frequency data within 1, 2 and 3 
   standard errors in those data reside in the inner, intermediate 
   and outer ellipses, respectively.          
  }
\label{f:errellipse}
\end{figure}
Another improvement is to remove from the diagnostics more of the influence of 
regions of the Sun that are outside the core. The absolute frequency of a low-degree
mode of oscillation feels almost all of the interior structure of the star in inverse
proportion to the sound speed \HL{along a ray path, except}  near the surface where the 
influence of the rapid variation of
the acoustical cutoff frequency $\omega_{\rm c}$ dominates. The latter is largely eliminated
in the small frequency separation, because the eigenfunctions in the very surface layers
are almost independent of $L$, and therefore subtracting two modes of nearly the same
frequency entails a high level of cancellation. However, the cancellation is not
complete, simply because the frequencies of the two modes are not exactly the same. As
\citet{ulrich86} has pointed out, the ratio $R_l$ of the small separation $d_l$ to the
large separation $\Delta_l$ is a more direct measure of age, for it isolates more
effectively the nonhomologous aspects of the evolution 
\HL{\citep{g90b}}, 
and it more effectively eliminates the influence of the outermost layers 
of the Sun, as can easily be appreciated by comparing the formulae for 
$d_l$ and $R_l$ implied by the asymptotic
expression~(\ref{e:asymp}). \citet{rv03} and \citet{fct05} 
have advocated that it be used for core calibration instead of $d_l$, 
and recently \citet{dbc11} have illustrated its robustness numerically. Here we have gone
further by adopting as diagnostics the factors $\hat A$, $\hat C$, and $\hat F$,
integrals of the solar structure which sense variations in
conditions even more concentrated towards the centre of the star.


\begin{figure}
\centering
\includegraphics[width=1.00\linewidth]{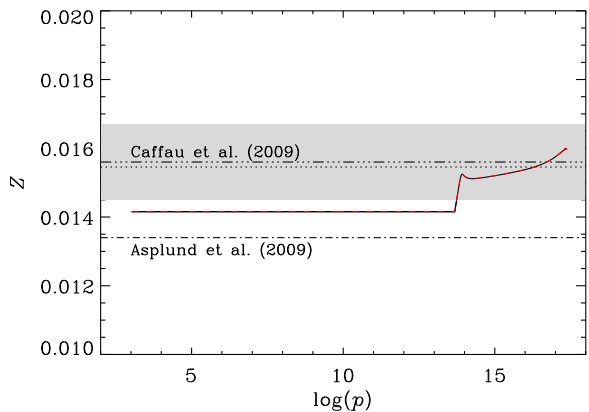}
\caption{
Heavy-element abundance $Z$ as a function of the depth-coordinate $\log(p)$ obtained from 
the reference Models~0 ($t_\star=4.60\,$Gy, solid black curve) and 2 
($t_\star=4.37\,$Gy, dashed red curve) after five calibration iterations.
Results from \HL{spectroscopic analyses based on} numerical simulations by \citet{asp09} (dot-dashed line) and
\citet{caf09} (triple-dot-dashed line) \HL{of the convectively driven macroscopic motion in 
the atmosphere} are indicated; the shaded area
indicates the reported error bars by \citet{caf09}. The initial (zero age)
value $Z_0$ of \HL{both reference models} 
is indicated by the dotted line. 
After five calibration iterations both reference models have a surface 
heavy-element abundance $Z_{\rm s}=0.0142$;  the age obtained from the 
4.60$\,$Gy reference model (Model~0) is 4.604$\pm0.039\,$Gy, and that from  
the 4.37$\,$Gy reference model (Model~2) is 4.603$\pm0.039\,$Gy
(see also Table~\ref{t:results}).
} 
\label{f:Zprofile}
\end{figure}


One could consider going even further by trying to replace 
\HL{the set of diagnostic factors with a single combination of $\hat A$, $\hat C$ and $\hat F$}
designed to eliminate the influence of the surface layers as much as possible, analogous  
to the procedure adopted  by \citet{gk88} and \citet{ketal92};
that is tantamount to using a 
\HL{judiciously selected combination of small separation ratios $R_l$}.
Because $\hat A$, $\hat C$ and $\hat F$ depend differently on the core stratification, the
simultaneous use of all three quantities provides some information about the manner in 
which $X$ varies with $r$. It is therefore to be hoped that the calibration is more secure
than one using just $d_l$ or $R_l$. It is worth mentioning at
this juncture that the 
\HL{integrand}
for $\hat A$ is not actually negligible outside the core, as can
be seen from Fig.~\ref{f:integrals}; indeed it has been known for some time that the
integrand continues to the surface with approximately the same magnitude as it has at
$r/R=0.5$ \HL{\citep[][see also Fig.\,\ref{f:AZ0}]{g86b}},
and that the integral is dominated by conditions outside the 
core. However, it appears that only the inner parts change as $t_\star$ and $Z_0$ vary,
and therefore that $\hat A$ is \HL{at least a fairly}  good diagnostic for our purposes. 
\HL{We note, however, that there is some contamination from outside the core, as is 
hinted in Table~\ref{t:calibration} in which it is recorded that $|C_{\Theta 1 2}|$ 
is smallest when $\hat A$ is not used in the calibration.}

It is also important to include the diagnostic $\hat\Gamma:=-\delta\gamma_1/\gamma_1$,
which measures the helium abundance $Y_{\rm s}$ in the convection zone, for that reflects
a rather different aspect of the core structure and 
\HL{thereby}
enables a much more precise
determination of $t_\odot$ and $Z_0$, as evinced by Table~\ref{t:calibration}, and
which was already evident in an earlier phase of the investigation \citep{hg07a}.
Whether or not the outcome is more accurate depends on the reliability of the procedure
to account for gravitational settling, which relates $Y_{\rm s}$ in the surface to \HL{the value of} 
$Y_0$ which controls conditions in the core. It should be pointed out also that $\hat\Gamma$
is not an uncontaminated measure of $Y_{\rm s}$, because it depends also on the entropy
in the deep adiabatically stratified convection zone \citep{hg07b}, \HL{ and perhaps in reality 
also on the existence of an intense magnetic field (see below).}  Our procedure could be
made more reliable if we could find an alternative diagnostic that senses $Y_{\rm s}$ more
directly.

Further remarks about the influence of the outer layers, or the elimination thereof, are 
in order: In fitting the resolvable glitch contribution to the data an 
approximation to the unresolvable contribution from hydrogen ionization and the upper
superadiabatically stratified boundary layer 
was included, equivalent to a cubic form in $\nu^{-1}$ added to the second differences
\citep{hg07b}. 
Associated with the resolvable glitches
are smooth contributions which were ignored in the initial calibration for
$t_\odot$ and $Z_0$ \citep{hg07a}. Subsequently they were taken explicitly into account,
thereby removing a bias in the procedure \citep{hg09a, hg09b} and, it is to be
supposed, improving the accuracy of the calibration. It should be mentioned, however, 
that we have not taken explicit account of putative errors in our modelling of the
outermost layers of the Sun. \citet{cdg80} found that were the oscillations to be
adiabatic, the effect of the atmosphere would  be to add to the frequencies 
a term $\delta$ that is
itself a rapid function of frequency: $\delta\propto\nu^b$ with $b=2(m+1)$ for
\HL{$\nu/\nu_{\rm c}\ll1$}, 
where $\nu_{\rm c}$ is the 
cyclic cutoff frequency \HL{$\omega_{\rm c}/2\pi$ and $m$ is an effective polytropic 
index in the vicinity of the
upper turning point, which, from fitting a (smooth) asymptotic frequency formula to 
solar data, is expected to have a value of about 3 \citep{g86a, hg07b};   
furthermore} ${\rm d}\ln\delta/{\rm d}\ln\nu$
decreases with increasing $\nu$ as $\nu/\nu_{\rm c}$ approaches and exceeds unity.
\citet{kbc08} found that in the Sun $b$ decreases to about 4.9 for 
\HL{$\nu\simeq3\,$mHz}, 
which is not entirely inconsistent with this finding.

\HL{
We note, furthermore, that the influence of the perturbations on the Reynolds stress
induced by the oscillations also has a component that increases with $\nu$,
but more slowly than the effect of a perturbation in the atmosphere \citep[e.g.][]{g86b, b92, rosen95, h10}.
This result may also be partially responsible for the exponent $b$ being somewhat less than
\HL{ the expected value of} 2$(m+1)$.
}
\citet{dbc11} found that taking the
correction into account obviates the necessity to use $R_l$ instead of $d_l$ in a
simple model calibration for $t_\odot$ in which $Z_0$ is held fixed, and yields results
similar to those obtained from $R_l$ with no surface term. This suggests that
our neglect of the near-surface adjustment -- a device which we adopted 
to maintain a workable number of
unknown parameters in the fitting -- may not be severely deleterious. Nonetheless, the 
approximation deserves further 
\HL{scrutiny}.

We have also been somewhat cavalier in our modelling of the acoustic glitches at the
base of the convection zone. In particular, we have modelled them as a simple discontinuity
in the second derivative of the density \HL{together with an exponential recovery beneath 
\citep{hg07b} to represent} 
standard solar models. Again, we have taken this approach for our convenience; after all, 
the sole purpose of modelling the glitch was to remove it. However, we are aware that we
have not adequately taken account of the stratification of the tachocline, and that by so
doing we risk not having eliminated adequately its contribution to the frequencies, and
thereby may have biassed our final result. Indeed, it is evident that we have not been
able to fit for the rapidly oscillating component of the second differences to the
solar data as well as we have to the frequencies of a standard solar model, suggesting
that there might be room for further improvement of the theory.
\citet{mt05} and \citet{cdmrt11} have gone some way in making such improvements, with 
the intention of studying the stratification at the base of the convection zone itself. It
behoves us to do so too.  \HL{In this regard we observe that the differences between 
the residuals from the Sun and the calibrated model, displayed in Fig.\,\ref{f:diff_residuals},
show evidence of  undersampled high-frequency oscillations that are $l$ dependent, hinting 
that the tachocline structure might be aspherical.} 
\begin{table*}
\centering
\caption{
Age calibrations for fixed initial heavy-element abundance $Z_0=0.019628$ 
(the value adopted for Model S; \citealt{jcd96}) and $Z_0=0.014864$ (the value 
adopted for a solar model assuming the \citealt{asp09} abundances; see also \citealt{jcdhg10}).
Results are shown for the combinations $\zeta_\alpha=(\hat A, \hat C, \hat F)$, 
$(\hat A, \hat C)$ and $(\hat A)$,
and for two reference models: Model\,0 with an age $t_\star=4.60\,$Gy 
and Model\,2 with an age $t_\star=4.37\,$Gy. The first three columns show the
calibrated ages and associated errors $\epsilon_\Theta$ for \HL{both} reference models
computed with constant $Z_0=0.019628$; the last three columns show the results 
for the age calibration \HL{of both} reference models computed with 
constant $Z_0=0.014864$. For the $Z_0=0.019628$ age calibration the reference 
models and derivatives 
$H_\alpha=({\rm d}\hat A/{\rm d}t, {\rm d}\hat C/{\rm d}t, {\rm d}\hat F/{\rm d}t)$ 
were obtained from a grid of 5 models with varying age adopting the \citet{greno93} 
solar composition.  The $Z_0=0.014864$ age calibration used the reference models 
and derivatives $H_\alpha$ from a grid of 5 models of varying age assuming 
the \citet{asp09} solar composition.
}
\begin{tabular}{lccccccc}
\noalign{\smallskip}
\noalign{\smallskip}
\hline
\noalign{\smallskip}
 &&$Z_0\,=\,$0.019628&\qquad&&&$Z_0\,=\,$0.014864&\\
 \hfil$\zeta_\alpha$\hfil&
 \hfil $t_\odot$ (Gy)\hfil&
 \hfil $t_\odot$ (Gy)\hfil&
 \hfil$\epsilon_\Theta$\hfil&
 $\qquad$&
 \hfil $t_\odot$ (Gy)\hfil&
 \hfil $t_\odot$ (Gy)\hfil&
 \hfil$\epsilon_\Theta$\hfil\\
 &(Model 0)&(Model 2)&\qquad&&(Model~0)&(Model 2)&\\
\hline
$\hat A,\hat C,\hat F    $&4.272&4.264&0.050&$\qquad$&4.414&4.408&0.054\\
$\hat A,\hat C           $&4.486&4.490&0.061&$\qquad$&4.585&4.587&0.061\\
$\hat A                  $&4.437&4.439&0.081&$\qquad$&4.559&4.561&0.081\\
\noalign{\smallskip}
\hline
\end{tabular}
\label{t:fixedZ}
\end{table*}

It is one of our intentions to refine our core diagnostic by combining the
integrals $\hat A$, $\hat C$ and $\hat F$ into a single quantity $\cal{T}$
which measures most closely the total hydrogen consumption $\Delta H$, rather 
than merely using the three different aspects of the deficiency function $h(r)$ in parallel. 
By so doing, properties
of the core that are not direct indicators of age should be partially eliminated,
thereby increasing the accuracy of the calibration; furthermore, the reduction of the 
number of final calibration parameters from four to two would increase 
the precision, although that is of secondary concern.
The construction of the diagnostic $\cal{T}$ is a tedious, although, we believe, relatively 
straightforward task which we have not yet completed.

Another of our unaccomplished intentions is to report on varying the bounding values
$k_1$ and $k_2$ of $k=n+\frac{1}{2}l$ between which the modes used in the calibration
are chosen to lie, as did \citet[][see also \citealt{hg08}]{dog01}. 
This should give a better 
indication of the robustness of the calibration. We have carried out a partial
survey, but we are not yet satisfied with the outcome. The reason is that
the function $E_{\rm g}$ defined by equation~(\ref{e:minsecdiff}), when evaluated with
the coefficients of a corresponding smooth model represented by the coefficients in the
expansion~(\ref{e:asymp}), has several local minima. The calibration we report here adopts the
lowest of those minima. But we have found that as $k_1$ and $k_2$ are varied the relative
depths of the minima change, and always selecting the lowest can lead to sudden jumping
from one to another. The situation is superficially not unlike the earliest direct solar model
calibration \citep{cdg81}, which also used only low-degree modes, and for which the
acceptable minimum had eventually to be determined from other, rather different, seismic data.
Maybe the resolution here will turn out to be similar.

The standard calibration errors quoted in Tables~\ref{t:calibration}--\ref{t:fixedZ} 
and illustrated in Fig.~\ref{f:errellipse} are the result of propagating quoted
observational errors in the raw frequencies. They indicate the precision of the
calibration. In the absence of information to the contrary, we have assumed that the 
raw-frequency errors
are uncorrelated. It is important to realize that, given that some correlation is
inevitable, this assumption can not only cause the
precision of the calibration to be overestimated, but can also lead to bias in the results
\citep{dog96, gs02}. The calibration errors evidently overestimate the precision.
And, of course, they certainly overestimate the accuracy.

Our calibration yields $Z_{\rm s}=$0.0142 
for the current surface heavy-element abundance of the Sun. This is significantly
smaller than that of Model S of \citet{jcd96}, which has almost the correct sound-speed
and density distribution throughout. Therefore our `best' model is wrong.  
\HL{This conclusion is borne out by the fact that the residual 
differences, plotted in Fig.\,\ref{f:diff_residuals}, are not zero.}  
\HL{In particular, the opacity in the Sun, which is what $Z$ principally determines, and 
which is fairly reliably determined by analysis of essentially all of the seismic modes
of intermediate and high degree \citep[][]{dog04}, is not faithfully reproduced by our fitted model.}
What does that imply about the values we infer for $Z_{\rm s}$ and $t_\odot$? 
\HL{Rather than calibrating their models for the heavy-element abundance,} 
others \citep[e.g.][]{wd99, bsp02, dbc11} have instead simply adopted a value
for $Z_0$ or $Z_{\rm s}$ that was perhaps acceptable by other 
criteria, and carried out a much more straightforward single-parameter calibration
to estimate $t_\odot$. The precision of such a calibration is greater than it would
have been had $Z_{\rm s}$ (or, equivalently, $Z_0$) been included as a fitting parameter,
but not necessarily the accuracy, even if the true value of $Z_0$ had been adopted. This
matter is discussed by \citet{g11}, who suggests that under conditions such as these,
a simple, but admittedly not reliable, rule of thumb is that accuracy tends to
decrease as precision increases. One cannot be sure that that is the case here without
a much deeper understanding of the properties of the models against which the Sun is
calibrated.

It is of some interest to record how the outcome of such single-parameter calibrations 
depend on the values assumed for $Z_0$. It is summarized in 
Table~\ref{t:fixedZ} for two constant heavy-element abundances: 
$Z_0$=0.019628, the value adopted for Model~S \citep[][]{jcd96}, 
and $Z_0$=0.014864, the value adopted for
the \citet{asp09} abundances \citep[see][]{jcdhg10}.
It is evident that a lower fixed value of $Z_0$ results in a greater solar age:  
\HL{an increase of 3\% associated with a 30\% decrease in $Z_0$. }
This is as one would expect. Reducing $Z_0$ requires also a reduction of $Y_0$ at
fixed age, resulting in a less centrally condensed star and consequently a greater
value of $\hat A$. Moreover, increasing the age at fixed $Z_0$ and $Y_0$ reduces
$\hat A$. Therefore, to maintain $\hat A$ constant, lowering $Z_0$ for the
calibration must be compensated by a rise in the inferred value for $t_\odot$.
It should be noted that the use of a value of $Z_0$ that is consistent with 
inferences from intermediate- and high-degree modes is a procedure which is
not available for calibrating stars other than the Sun.

Of course the reliability of the results of a calibration can be no greater than the 
reliability of the models that are used. Thus one should address the validity of the
assumptions that are made, and estimate their influence on the inferred values of
$t_\odot$ and $Z_0$. For making the estimate we note that
the final calibration is based on the values of the
parameters $\zeta_\alpha$: the coefficients $\hat A$, $\hat C$ and $\hat F$ 
of the most $L$-sensitive terms at each order in the asymptotic expression\,(\ref{e:asymp}),
and the measure $\hat\Gamma$ of the depression in $\gamma_1$ due to He$\,$II ionization.
Were the Sun to be spherically symmetrical and nonmagnetic, the former 
\HL{would be}
indicators of 
the acoustic stratification of the core, and the latter a (model-dependent) measure of
the surface helium abundance which is related, via the solar model, to the helium 
abundance in the core in a weakly $t_\odot$- and $Z$-dependent way. 
Here we address the potential errors in our estimates of the values of these two quantities.
For illustrative purposes we lump the first three parameters together, and consider
only $\zeta_1=\hat A$, together with $\zeta_4=\hat\Gamma$. Then, from the derivatives listed
in Table~\ref{t:hij} one can deduce that for small errors $\delta\hat A$, 
$\delta\hat\Gamma$ in $\hat A$ and $\hat\Gamma$ the corresponding errors in 
$t_\odot$ and $Z_0$ are determined by 
\begin{equation}
\left(
\begin{array}{c}
\delta\ln t_\odot \\
\delta\ln Z_0
\end{array}
\right)
=
\left(
\begin{array}{cc}
-0.91  & -0.58 \\
-3.2& 3.8 
\end{array}
\right)
\left(
\begin{array}{c}
\delta\ln\hat A  \\
\delta\ln\hat\Gamma
\end{array}
\right)\,.
\label{e:caliberr}
\end{equation}

The value of $\hat A$ obtained by fitting the expression~(\ref{e:asymp}) to the `smooth'
frequencies can be misinterpreted by ignoring asphericity, which arises principally from
solar activity in the superficial layers of the Sun. 
\HL{
Formally, for spherically symmetric stars, a change in $L$ 
at fixed  seismic frequency $\nu$ is associated with a change in 
the depth of the lower turning point, which is what we use to gauge the modification
of the stratification resulting from hydrogen burning. But there is also a modification of the 
$L$ dependence by asphericity, which we have not taken into account here.
This comes about because of azimuthal-order-dependent  filtering
produced by whole-disc measurements does not produce mean multiplet frequencies  but
is biassed towards the sectoral modes. Here we have taken the \citet{basu07} corrections based
on a broad single-proxy correlation, rather than having tried to estimate the bias from the actual
latitudinal distribution of solar activity. 
However,  we might look at scatter and estimate the residual.
To estimate the effect on our calibration we use the observations of
}
\citet{cemv07}, who plot mean frequency differences at different epochs, averaged over
$n$, for different values of degree $l$. These can be fitted to $L^2$ to estimate the
corruption $\delta_{\rm a}\hat A$ to the coefficient $\hat A$. Averaging over the interval of
observations of the \citet{basu07} data set that we use for our calibration yields 
\HL{$\overline{\delta_{\rm a}\hat A}\simeq-5\times10^{-3}$}, 
implying that $t_\odot$ would be 
\HL{overestimated} 
by 0.06$\,$Gy and $Z_0$ overestimated by 0.009.
These systematic changes are not negligible: the change in $t_\odot$ is comparable
with, although somewhat larger than, the typical random errors listed in
Tables~\ref{t:calibration} and \ref{t:results} in calibrations that use $\hat\Gamma$;
the change in $Z_0$ is 
\HL{some twenty}
times larger. Asphericity arising from the
centrifugal force of rotation is negligible for the Sun, but it can be
significant in rapidly rotating stars. The asphericity of the solar tachocline
is also insignificant at our present level of precision. 

\begin{table*}
\centering
\caption{\HL{Age $t_\odot$ and \HL{initial heavy-element abundance $Z_0$} of 
calibrated solar models obtained after five iterations from reference 
models {Model~0} and {Model~2} using the parameter 
combinations $\zeta_\alpha=(\hat A, \hat C, \hat F, -\delta\gamma_1/\gamma_1)$ 
and $(\hat C, \hat F, -\delta\gamma_1/\gamma_1)$. }
\HL{The values for the initial helium abundance $Y_0$ and surface abundances 
$Z_{\rm s}$ and $Y_{\rm s}$ are obtained from the models.}
The last \HL{three} columns \HL{are} the standard errors 
(components of the error covariance matrix C$_{\Theta}$) 
\HL{associated with the calibrated values of $t_\odot$ and $Z_0$}.
}
\begin{tabular}{lccccccccc}
\noalign{\smallskip}
\noalign{\smallskip}
\hline
 \hfil$\zeta_\alpha$\hfil&
 \hfil $t_\odot$ (Gy)\hfil&
 \hfil$Z_{\rm 0}$\hfil&
 \hfil$Y_{\rm 0}$\hfil&
 \hfil$Z_{\rm s}$\hfil&
 \hfil$Y_{\rm s}$\hfil&
 \hfil C$^{1/2}_{\Theta 11}$\hfil&
 \hfil -(-C$_{\Theta 12})^{1/2}$\hfil&
 \hfil C$^{1/2}_{\Theta 22}$\hfil\\
\hline
\noalign{\smallskip}
& \multicolumn{7}{c}{Model 0}\\
\noalign{\smallskip}
$A,C,F,-\delta\gamma_1/\gamma_1$&4.604&0.0155&0.250&0.0142&0.224&0.039&0.0013&0.0005\\
$C,F,-\delta\gamma_1/\gamma_1  $&4.602&0.0155&0.251&0.0142&0.224&0.044&0.0004&0.0005\\
\hline
\noalign{\smallskip}
& \multicolumn{7}{|c|}{Model 2}\\
\noalign{\smallskip}
$A,C,F,-\delta\gamma_1/\gamma_1$&4.603&0.0155&0.250&0.0142&0.224&0.039&0.0013&0.0005\\
$C,F,-\delta\gamma_1/\gamma_1  $&4.601&0.0155&0.251&0.0142&0.224&0.044&0.0004&0.0005\\
\noalign{\smallskip}
\hline
\end{tabular}
\label{t:results}
\end{table*}

Errors in the diagnostics of $Y_0$ have two obvious main sources. The first is the 
relation between $\delta\gamma_1$ and $Y_{\rm s}$, which depends on the equation
of state, which we know is not accurate to a degree of precision that we would like.
The matter has been discussed extensively by \citet{ketal92},
\citet{cdd92} and by \citet{betal00}, and we do not pursue it here.
The second is the relation between $Y_{\rm s}$ and $Y_0$, and also $Z_{\rm s}$ and $Z_0$,
which depend on gravitational settling. It is difficult to assess the accuracy of
currently used prescriptions, and it is likely that the uncertainty will remain
with us for some time. Yet we note that it is not unlikely that the uncertainty
exceeds our statistical errors arising from data errors.
We note first that $Y_0$ and 
\HL{$Y_{\rm s}$ differ by about 0.026}. 
This represents
the amount of gravitational settling out of the convection zone that has taken 
place over the lifetime of the Sun. If the computation of the settling rate were
underestimated by 20\%, say \HL{--  an error which we do not regard as being unrealistically 
high, given the uncertainty in the value of the Coulomb logarithm used in truncating the 
electrostatic particle interactions \citep[cf.][]{michproff93} -- }
then the effective $\delta\hat\Gamma$ 
\HL{could have been}
underestimated likewise, and according to equation\,(\ref{e:caliberr})
errors of 
\HL{+0.05\,Gy}
and -0.001 would be imparted to
$t_\odot$ and $Z_0$ respectively. 
What is perhaps more serious is the possibility of material redistribution
in the energy-generating core, either by large-scale convection or by
small-scale turbulence induced possibly by rotational shear. \citet{gk90} 
argued that there is evidence for that having occurred, concomitant with a
reduction of the sound-speed gradient in the innermost regions, and
thereby making the Sun appear younger than it really is.  
This matter should perhaps be investigated further in the future.
But what requires serious consideration now is the degree to which a magnetic
field might suppress the acoustic glitch associated with helium ionization.
\citet{bm04} and \citet{vce06} have reported a diminution during solar
cycles~22 and 23 in the amplitude of the acoustic signature of the glitch
with increasing magnetic activity (gauged by the 10.7$\,$cm radio flux $F_{10.7}$),
with an average slope ${\rm d}\ln\hat\Gamma/{\rm d}F_{10.7}\simeq-0.001$
(in units of 10$^{22}$J$^{-1}$s$\,$m$^2\,$Hz).   It has already been pointed out that that
requires magnetic field strength variations of order 10$^5$G 
\HL{in the second helium ionization zone} \citep{dog06}.
Moreover, it is much greater than that implied by Libbrecht \& Woodard's (1990) 
observations in the previous cycle.
Given that the average of $F_{10.7}$ over the interval of observation of the BiSON
data was about 120, this would imply, had the magnetic perturbations been small, that
$\hat\Gamma$ has been underestimated by about 10\%, namely 7$\times$10$^{-3}$, implying
that $t_\odot$ has been overestimated by about 10\% and, formally, $Z_0$ underestimated
by about 90\%. This result appears to render hopeless any attempt to calibrate the
glitch to determine $Y_{\rm s}$. \HL{However, we note that recently \cite{cdmrt11} 
have found no evidence for such variation.} It behoves us, therefore, urgently to investigate the 
matter further.  Magnetic-field issues aside, the relation between $\hat\Gamma$ and $Y_{\rm s}$ is 
\HL{reliant}
on the equation of state, which we know to be deficient \citep[e.g.][]{ketal92, betal00}.

There are other assumptions that are implicit in most solar evolution calculations.
Two which have obvious serious implications regarding the apparent age are the
constancy of the total mass $M$ 
of the Sun -- the assumption is that there has been
no significant accretion nor mass loss on the main sequence -- and that physics has
not evolved such that, in appropriate units, Newton's gravitational constant $G$ 
varies with time. Failure of either of those two assumptions can lead to a
substantial deviation of the Sun's evolution from the usual standard.
Numerical computations of the effect of varying $G$ were carried out long ago
by \citet{ps64}, \citet{ec65}, \citet{rd66} and \citet{sb69}, and computations with
mass loss have been performed by \citet{gu87}, 
\HL{\citet{sf92}, \citet{gc95}, \citet{sb03} and \citet{gumu10}}; the results 
\HL{are summarized  by an analytical approximation given by} 
\HL{\citet{g90b}}.
In particular, if $G$ or 
\HL{$M$}
had been greater in the past, then the solar luminosity would
have been greater, and more hydrogen would have been consumed, \HL{and the Sun would 
now be younger than it appears.  It behoves us also to mention that the luminosity of the Sun 
today is somewhat greater than the standard value that we have adopted in this work, 
because the surface radiant flux increases slightly with increasing latitudinal direction.}
Such issues go beyond the scope of this investigation.

\section{Conclusion}

We have attempted a seismic calibration of \HL{standard} solar models with a view to 
improving earlier estimates of the main-sequence age $t_\odot$ and the initial 
heavy-element abundance $Z_0$.  
\HL{Our long-term goal has been to achieve a precision which could distinguish 
between planet formation occurring simultaneously with or subsequent to the formation of the Sun.} 
Our current best estimates\HL{, around $4.60\pm0.04$Gy,}  are summarized in Table\,\ref{t:results}: 
the age is close to the previous preferred values -- in particular, the age adopted for
Christensen-Dalsgaard's Model S -- and the implied present-day surface heavy-element
abundance lies between the modern spectroscopic values quoted by \citet{asp09} and
\citet{caf09}. 
However, we emphasize that there remain many uncertainties in our procedure, and that
future revision is not unlikely.


  \subsection*{ACKNOWLEDGEMENTS}

We are grateful to Bill Chaplin for supplying the BiSON 
data plotted in Fig.\,\ref{f:BiSON}. Support by the Austrian 
FWF Project P21205-N16 is gratefully acknowledged.
DOG is grateful to the Leverhulme Trust for an Emeritus Fellowship.



\end{document}